\documentclass[]{aa}

\usepackage[varg]{txfonts}
\usepackage{graphicx}
\usepackage{multicol}
\usepackage{amsmath}
\usepackage[colorlinks=true,linkcolor=magenta,citecolor=blue,filecolor=cyan,urlcolor=black,final=true]{hyperref}
\usepackage[normalem]{ulem}
\usepackage{xcolor}

\begin{document} 

\makeatletter 

\renewcommand*\aa@pageof{, page \thepage{} of \pageref*{LastPage}}
\makeatother

   \title{Multi-spacecraft observations of the structure of the sheath of an interplanetary coronal mass ejection and related energetic ion enhancement}
	\titlerunning{ICME sheath structure and ion enhancements}

   \author{E.~K.~J.~Kilpua \inst{1}
        \and
       S.~W.~Good \inst{1}
        \and
       N.~Dresing \inst{2,3}
       \and
       R.~Vainio \inst{2}
       \and
       E.~E.~Davies \inst{4}
       \and
       R.~J.~Forsyth \inst{4}
       \and
       J.~Gieseler \inst{2}
       \and
       B.~Lavraud \inst{5,6}
       \and
       E.~Asvestari \inst{1}
       \and
       D.~E.~Morosan \inst{1}
        \and
       J.~Pomoell \inst{1}
       \and
       D.~J.~Price \inst{1}
       \and
       D.~Heyner \inst{7}
       \and
       T.~S.~Horbury \inst{4}
       \and
       V.~Angelini \inst{4}
       \and
       H.~O'Brien \inst{4}
       \and
       V.~Evans \inst{4}
        \and
       J.~Rodriguez-Pacheco \inst{8}
       \and
       R.~Gómez Herrero  \inst{8}
       \and
       G.~Ho \inst{9}
       \and
       R.~Wimmer-Schweingruber \inst{3}
      }

   \institute{Department of Physics, University of Helsinki, P.O. Box 64, FI-00014 Helsinki, Finland \\
              \email{emilia.kilpua@helsinki.fi}
         \and
           Department of Physics and Astronomy, FI-20014 University of Turku, Finland
           \and
           IEAP, University of Kiel, Germany 
           \and
            Department of Physics, Imperial College London, London, UK
           \and
           Laboratoire d'Astrophysique de Bordeaux, Univ. Bordeaux, CNRS, Pessac, France
           \and
           IRAP, CNRS, UPS, CNES, Université de Toulouse, Toulouse, France
           \and
           Technical University of Braunschweig, Braunschweig, Germany
           \and
          Space Research Group, Universidad de Alcalá, Madrid, Spain
           \and 
           Johns Hopkins Applied Physics Laboratory, Laurel, USA
           \\
             }
           
   \date{Received 19 March 2021 / Accepted 8 April 2021}

 
  \abstract
    {Sheath regions ahead of coronal mass ejections (CMEs) are large-scale heliospheric structures that form gradually with CME expansion and propagation from the Sun. Turbulent and compressed sheaths could contribute to the acceleration of charged particles in the corona and in interplanetary space, but the relation of their internal structure to the particle energization process is still a relatively little studied subject. In particular, the role of sheaths in accelerating particles when the shock Mach number is low is a significant open research problem.}
    {This work seeks to provide new insights on the internal structure of CME-driven sheaths with regard to energetic particle enhancements. A good opportunity to achieve this aim was provided by multi-point, in-situ observations of a sheath region made by radially aligned spacecraft at 0.8 and $\sim 1$ AU (Solar Orbiter, the L1 spacecraft Wind and ACE, and BepiColombo) on April 19-21, 2020. The sheath was preceded by a weak and slowly propagating fast-mode shock.}
    {We apply a range of analysis techniques to in-situ magnetic field, plasma and particle observations. The study focuses on smaller scale sheath structures and magnetic field fluctuations that coincide with energetic ion enhancements.}
    {Energetic ion enhancements were identified in the sheath, but at different locations within the sheath structure at Solar Orbiter and L1. Magnetic fluctuation amplitudes at inertial-range scales increased in the sheath relative to the solar wind upstream of the shock, as is typically observed. However, when normalised to the local mean field, fluctuation amplitudes did not increase significantly; magnetic compressibility of fluctuation also did not increase within the sheath. Various substructures were found to be embedded within the sheath at the different spacecraft, including multiple heliospheric current sheet (HCS) crossings and a small-scale flux rope. At L1, the ion flux enhancement was associated with the HCS crossings, while at Solar Orbiter, the ion enhancement occurred within a compressed, small-scale flux rope.}
    {Several internal smaller-scale substructures and clear difference in their occurrence and properties between the used spacecraft was identified within the analyzed CME-driven sheath. These substructures are favourable locations for the energization of charged particles in interplanetary space. In particular, substructures that are swept from the upstream solar wind and compressed into the sheath can act as effective acceleration sites. A possible acceleration mechanism is betatron acceleration associated with a small-scale flux rope and warped HCS compressed in the sheath, while the contribution of shock acceleration to the latter cannot be excluded.}


\maketitle


\section{Introduction}

{Coronal mass ejections \citep[CMEs; e.g.,][]{Webb2012} are huge eruptions of plasma and magnetic field from the Sun. After launch, CMEs expand and propagate through the heliosphere, where they are commonly called interplanetary CMEs (ICMEs). When their speed is sufficiently larger than that of the preceding solar wind, a collisionless fast shock and a sheath region form ahead of the ICME \citep[e.g.,][]{Kilpua2017}. The compressed nature of sheaths, with their embedding of large-amplitude magnetic field fluctuations, makes them important drivers of a wide variety of space weather effects at Earth, including geomagnetic storms, strong auroral disturbances and dramatic changes in outer radiation belt electron fluxes  \citep[e.g.,][and references therein]{Kilpua2017SSR}.}

{Another key space weather feature in which sheaths likely play a role \citep[e.g.,][]{Manchester2005} is the acceleration of charged particles in the corona and in interplanetary space.  Fast CMEs and flares are regularly associated with solar energetic particle (SEP) events in which the particles can attain energies up to a few hundred MeV and occasionally up to a few GeV \citep[e.g.,][]{Reames2013}. ICMEs can drive shocks at large distances from the Sun and accelerate charged particles in a continuous manner \citep[e.g.,][]{Giacalone2012}. CME-related SEPs are characterised by long-lasting proton enhancements traditionally called 'gradual SEP events', while brief, electron-rich, flare induced SEP events are termed 'impulsive' \citep[][and references therein]{Reames2013}. SEP events are of major interest in solar-terrestrial studies due to their potential for causing hazardous space weather effects \citep[e.g.,][]{Vainio2009}. In particular, they are a concern for satellites in orbit and for astronaut safety, and they can also affect the composition and dynamics of the upper atmosphere \citep[e.g.,][]{Krivolutsky2012,Desai2016}. The largest SEP intensities at Earth are associated with gradual, CME-related events that sometimes span a large range of longitudes ($> 100^{\circ}$) in the heliosphere \citep[][]{Dresing2012,Desai2016}.}

{There are many open questions related to the physical mechanisms and efficiency of the acceleration process operating at CME-driven shock waves. One puzzle is why a significant fraction of SEP events are associated with CMEs that have low Mach number shocks in the corona \citep[e.g.,][]{Kouloumvakos2019}. Recently \cite{Giacalone2020} related energetic ions with the 2018 November 11 ICME detected by Parker Solar Probe (PSP) at about 0.25 AU from the Sun. The characteristics of this SEP event suggested that the acceleration occurred closer to the Sun by the weak shock that had dissipated before reaching $\sim 0.25$ AU. In high Mach number shocks, diffusive shock acceleration (DSA) is regarded as the dominant cause of energization. In this process, charged particles reflect back and forth from magnetic field irregularities around the shock \citep[e.g.,][]{Axford1977,Bell1978} and the process can be enhanced in the solar corona and interplanetary space significantly by streaming instabilities driving Alfv\'en waves in the the foreshock region to become unstable \citep[e.g.,][]{Lee1983,Vainio2007,Vainio2008}. However, DSA is unlikely to be the main cause of acceleration for low Mach number shocks because the wave-generation process strongly depends on the Mach number of the shock \citep{Afanasiev2015,Afanasiev2018}. }

{In particular in the case of low Mach number shocks the CME sheath region could play a significant role.  Enhanced levels of turbulence and compression \citep{Moissard2019,Good2020,Kilpua2020} immediately downstream of shocks have been invoked as modulating influences on the acceleration \citep[e.g.,][]{Lario2019}, but the whole sheath could contribute to the energization. The sheath regions also exhibit an abundance of waves and substructures. Processes occurring at the shock layer and near leading edge of the ejecta  \citep{Comas1988} provide free energy for the generation of various plasma waves, such as mirror modes and Alfvén ion cyclotron waves \citep[][]{Alalahti2018,Alalahti2019}. Sheaths also contain a wealth of substructures that either form in the sheath, or that are pre-existing structures swept into and compressed by the sheath. Examples of substructure found in sheaths include heliospheric current sheet (HCS) crossings \citep[e.g.,][]{Crooker1993,Neugebauer1993}, high-density piled-up compression regions and plasma depletion layers \citep[][]{Das2011}, magnetic reconnection exhausts \citep[e.g.,][]{Feng2013} and small-scale flux ropes \citep[][]{Kilpua2020}. Both the internal structure of CME sheaths and its relation to particle energization, however, are still relatively little studied.}

{In this paper, we examine a CME sheath and associated particle acceleration using observations from the recently launched Solar Orbiter and BepiColombo spacecraft, as well as from Wind and ACE at L1. At the time of the studied events, Solar Orbiter was located $\sim 0.8$ AU away from the Sun and BepiColombo close to L1. All spacecraft were almost radially aligned. The CME was observed in the solar wind during 19--21 April, 2020 (Davies et al., 2021) and was preceded by a slow-speed and weak interplanetary shock. Our study focuses on the smaller scale structures and magnetic field fluctuations found within the sheath, the differences in these structures and fluctuations at the various spacecraft, and the relation of the sheath structure to energetic ion enhancements. The paper is organised as follows: In Section 2, data sets are described. In Section 3, results are presented; an overview of the solar wind plasma and energetic particle observations during the event is given, followed by analysis of the fluctuations and smaller-scale sheath structures, with the focus being on intervals when energetic ion enhancements were observed. The results are discussed in Section 4, and conclusions are presented in Section 5.}


\section{Observations and data analysis} \label{sec:analysis}

\subsection{Data sets and spacecraft locations} 

Solar wind data sets from the Solar Orbiter \citep[SolO][]{Muller2013}, BepiColombo \citep{Benkhoff2010}, Wind \citep{Ogilvie1997} and Advanced Composition Explorer \citep[ACE;][]{Stone1998} spacecraft have been used. Solar wind plasma observations were available from L1 only. From Solar Orbiter, particle observations from the Energetic Particle Detector \citep[EPD;][]{Rodriguez2020} and magnetic field measurements from the magnetometer \citep[MAG;][]{Horbury2020} have been used. BepiColombo magnetic field data comes from the magnetometer on board the Mercury Planetary Orbiter \citep[MPO;][]{Glassmeier2010}. From Wind, magnetic field data from the Magnetic Fields Investigation (MFI) magnetometer \citep{Lepping1995}, plasma data from Solar Wind Experiment \citep[SWE;][]{Ogilvie1995}, and suprathermal electron observations from the Three-Dimensional Plasma and Energetic Particle Investigation \citep[3DP;][]{Lin1995} have been used. ACE data comes from the Magnetometer \citep[MAG;][]{Smith1998} and Electron, Proton and Alpha Monitor \citep[EPAM;][]{Gold1998} instruments.

\begin{figure}[ht]
\centering
\includegraphics[width=0.5\linewidth]{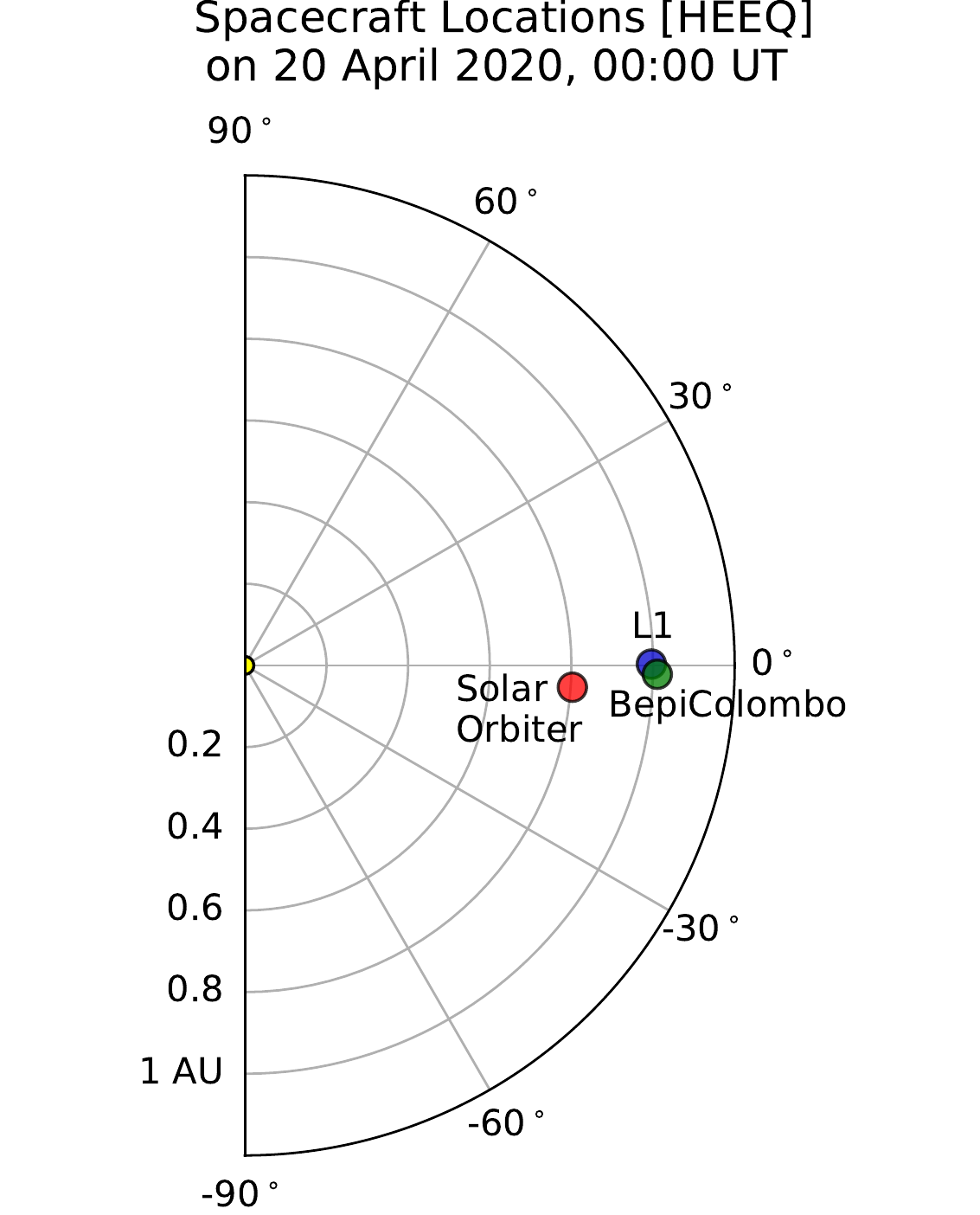}
\includegraphics[width=0.49\linewidth]{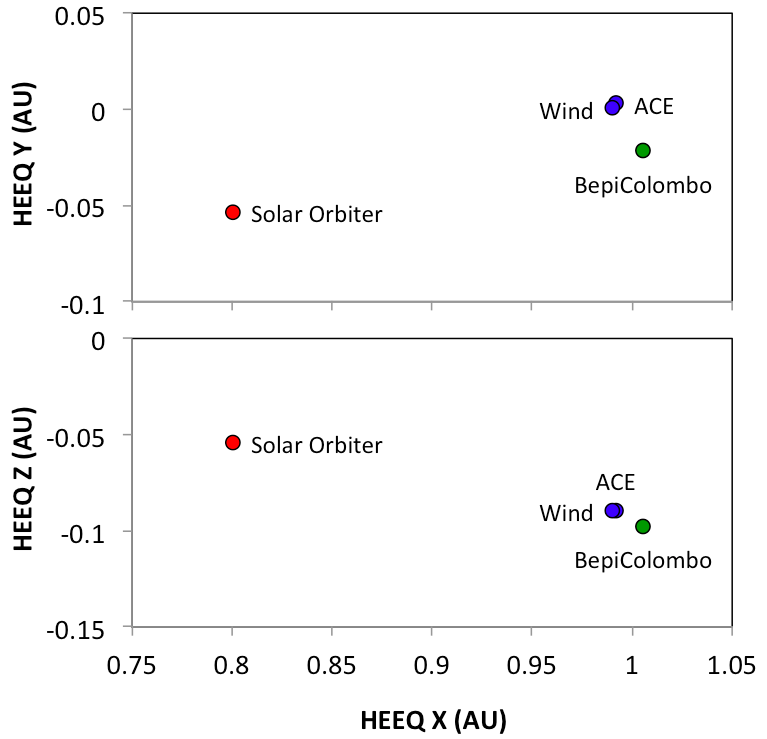}
\caption{Locations of Solar Orbiter, BepiColombo and the L1 spacecraft on 20 April, 2020 at 00:00~UT.}
\label{fig:location}
\end{figure}

The locations of Wind, ACE, BepiColombo and Solar Orbiter on April 20, 2020 $\sim~$00:00~UT are shown in Figure \ref{fig:location}. The spacecraft locations in Heliocentric Earth Equatorial (HEEQ) coordinates are given in Table \ref{tab:location}. Wind and ACE were both at the Lagrangian point L1, very close to 1~AU and along the Sun--Earth line. BepiColombo was also located in the vicinity of Earth, eastward of the Sun--Earth line. Solar Orbiter was closer to the Sun, at a heliocentric distance of $\sim 0.80$ AU, but also relatively close to the Sun-Earth line. In the HEEQ $Y$-direction, Wind and BepiColombo were separated by 587 $\mathrm{R}_\mathrm{E}$ (0.025 AU) and BepiColombo and Solar Orbiter by 916 $\mathrm{R}_\mathrm{E}$ (0.039 AU). 

\begin{table}
\caption{Spacecraft location at 0 UT on 20 April, 2020 in HEEQ coordinates.}
\centering
\begin{tabular}{l l l l l}
\hline
Spacecraft & X [AU] & Y [AU]   & Z [AU]   & r [AU] \\
\hline
\hline
Solar Orbiter & 0.80 & -0.053  & -0.054  & 0.80 \\ 
Wind (L1)     & 0.99 & 0.0032  & -0.090   & 1.00 \\
ACE (L1)     & 0.99 & 0.00068 & -0.090   & 0.99 \\
BepiColombo   & 1.01 & -0.022 & -0.098    & 1.01 \\
\end{tabular}
\label{tab:location}
\end{table}

\section{Results} \label{sec:results}

\subsection{Overview of the solar wind observations}

\begin{figure*}[ht]
\centering
\includegraphics[width=0.99\linewidth]{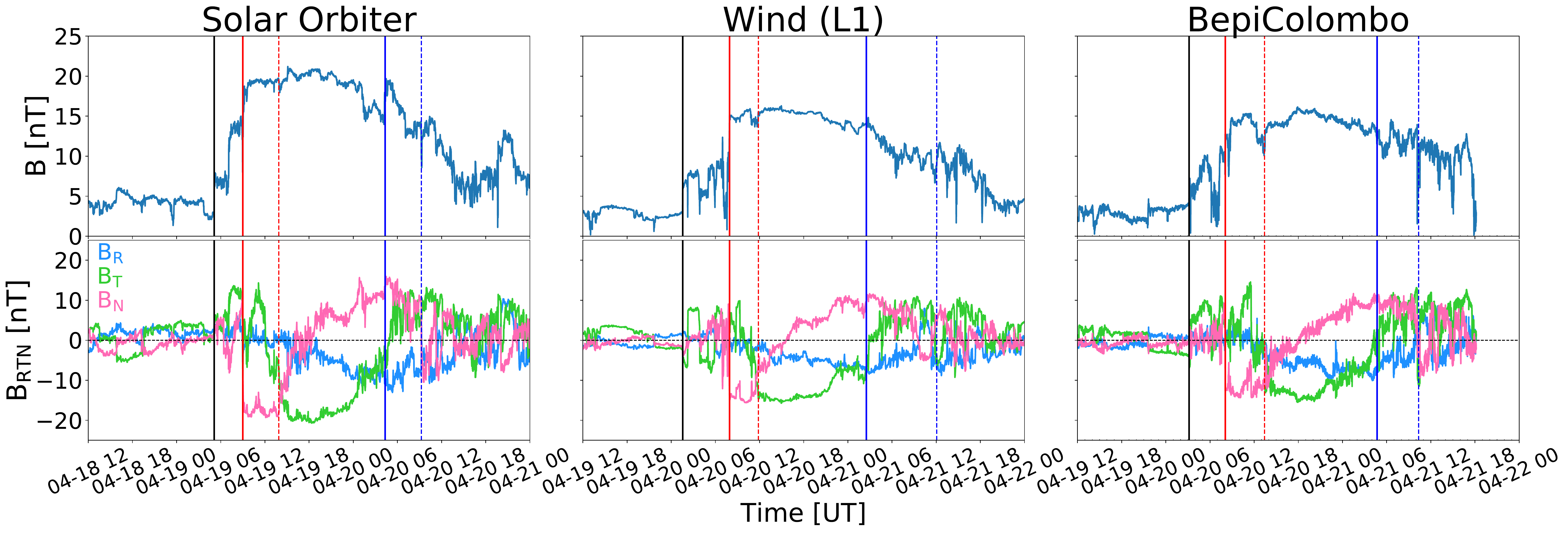}
\caption{Interplanetary magnetic field magnitude (upper panels) and components in RTN coordinates (lower panels) at Solar Orbiter, Wind and BepiColombo. Black vertical lines mark the shock. Solid and dashed red (blue) lines indicate the leading (trailing) edge times LE1 and LE2 (TE1 and TE2), respectively (see Table 2 and text for more details).}
\label{fig:FIG_Overview}
\end{figure*}

\begin{figure}[ht]
\centering
\includegraphics[width=0.99\linewidth]{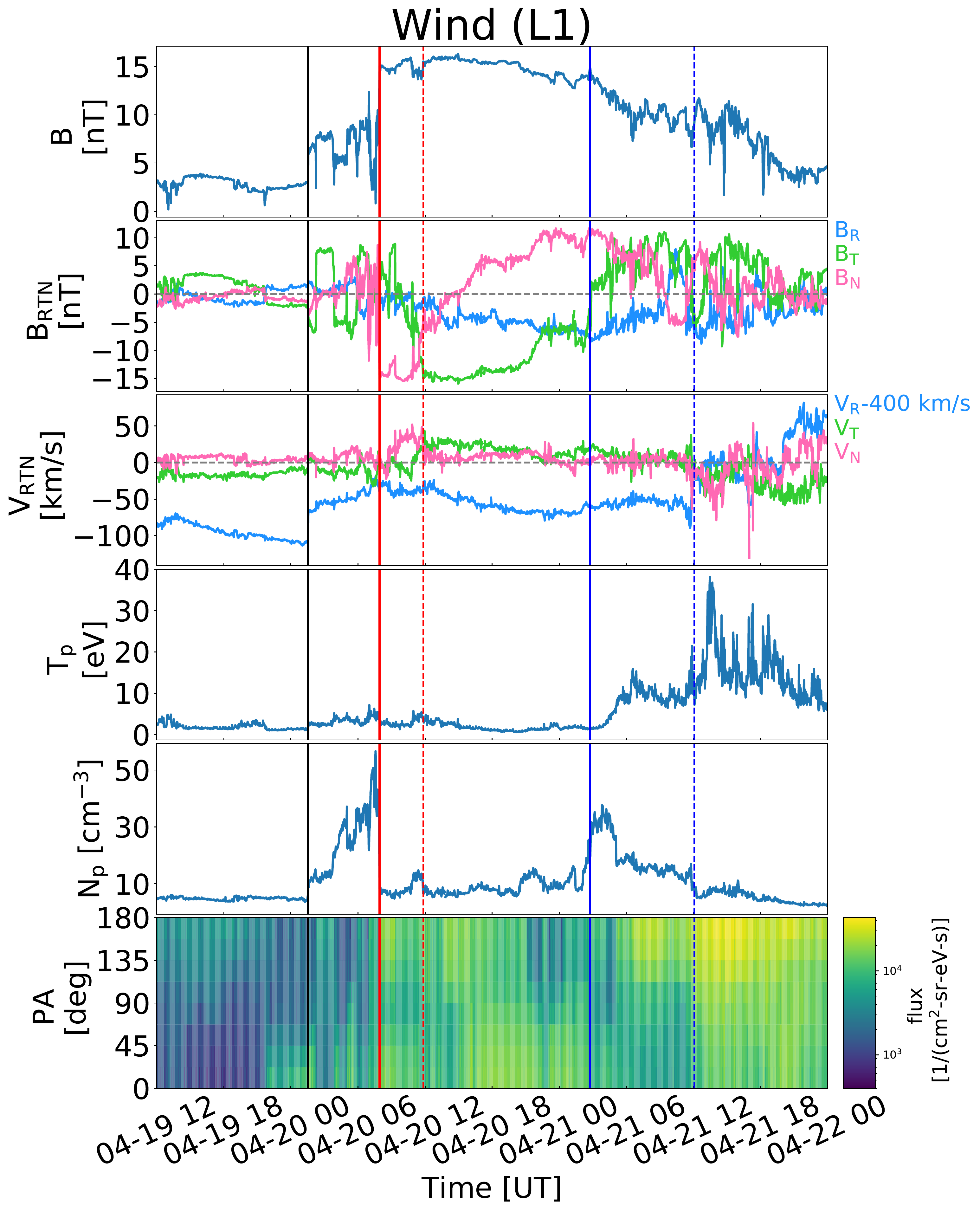}
\caption{Wind magnetic field and plasma data. From top to bottom, the panels show the magnetic field magnitude, magnetic field components in RTN coordinates, solar wind speed in RTN coordinates, proton temperature, proton density, and pitch angle distribution of suprathermal electrons at 265 eV. Vertical lines are the same as in Figure \ref{fig:FIG_Overview}.}
\label{fig:FIG_Overview_WIND}
\end{figure}

{An overview of the solar wind observations during a 3-day period centred on the ICME event at the three spacecraft is now presented. At all spacecraft, standard ICME signatures -- a leading shock, sheath and well-defined ejecta -- were observed. The magnetic field signatures were qualitatively similar at the three spacecraft, as can be seen in Figure~\ref{fig:FIG_Overview}. Plasma observations at Wind, including the bulk solar wind velocity components in RTN coordinates ($V_R - 400$ km/s), proton density and proton temperature are shown in Figure~\ref{fig:FIG_Overview_WIND}. Also shown in the figure is the colour-coded pitch-angle (PA) distribution of 265 eV suprathermal electrons at Wind. Suprathermal electrons are useful for investigating the magnetic connectivity to the Sun and magnetic topology of solar wind structures \citep[e.g.,][]{Gosling1987a}. Unidirectional heat flux flow is generally interpreted as an indication that magnetic field lines are connected to the Sun at only one end, while counter-streaming beams both parallel and anti-parallel to the field indicate closed magnetic structures (e.g., magnetic clouds) or suprathermal electron leakage from shocks or compressed solar wind structures \citep[e.g.,][]{Lavraud2010}. The times of the shock and the ejecta boundaries are given in Table \ref{tab:times}. There is ambiguity in both the leading and trailing edge times of this ejecta, and two possible times for these boundaries are given in Table \ref{tab:times}. This ambiguity is discussed further below.}

\begin{table*}
\caption{Shock, magnetic cloud and small-scale flux rope times and selected properties at Solar Orbiter, Wind and BepiColombo. The shock time, shock normals ($\textbf{n}_{\mathrm{sh}}$) as calculated from the magnetic coplanarity method, shock angles ($\theta_{\mathrm{Bn}}$ ), and downstream-to-upstream magnetic field ratios ($B_{\mathrm{d}}/B_{\mathrm{u}}$) are listed. The durations of the upstream/downstream averaging windows used in the determination of these parameters are: Solar Orbiter 5 minutes/4 minutes; Wind 10 minutes/10 minutes; and BepiColombo 8 minutes/4 minutes. The sheath duration ($\Delta T_{\mathrm{sheath}}$) is also given. For the magnetic cloud, the leading edge (LE) and trailing edge (TE) times are listed. The cloud boundaries were ambiguous for this event, and so two possible leading and trailing edge times are given here. For the small-scale flux rope (SFR), the table lists: the leading and trailing edge times; SFR duration ($\Delta T_{\mathrm{SFR}}$ ); mean magnetic field magnitude ($\langle \mathrm{B}_{\mathrm{SFR}} \rangle$); the latitude ($\theta_{\mathrm{SFR}}$) and longitude ($\phi_{\mathrm{SFR}}$) of the central axis direction; and the impact parameter ($p$), i.e., the ratio of the closest approach distance of the spacecraft from the SFR axis to the SFR radius.}
\centering
\begin{tabular}{l l l l}
\hline
Spacecraft & Solar Orbiter & Wind (L1) & BepiColombo   \\
\hline
\hline
Shock/Sheath      &  & & \\
\hline
Shock time [UT]          & 4/19 05:07 &  4/20 01:33 & 4/20 03:09 \\
$\textbf{n}_{\mathrm{sh}}$ & [0.97, -0.23, -0.0057] & [0.92,  -0.025, 0.38] & [0.55, -0.58, -0.59] \\
$\theta_{\mathrm{Bn}}$ & $43^{\circ}$ & $73^{\circ}$ & $50^{\circ}$  \\
$B_{\mathrm{d}}/B_{\mathrm{u}}$  & 2.3 & 2.1 & 1.7 \\
$\Delta T_{\mathrm{sheath}}$ [h] & 3.9    &  6.4        & 4.6   \\ 
\hline
Magnetic cloud & & & \\
\hline
LE1 [UT]        & 4/19 08:59 &  4/20 07:55 &  4/20 08:05 \\
LE2 [UT]        & 4/19 13:53 &  4/20 11:50 & 4/20 13:25  \\
TE1 [UT]        & 4/20 04:20 &  4/21 02:45 &  4/21 04:42 \\
TE2 [UT]        & 4/20 09:15 &  4/21 12:05 & 4/21 10:20 \\
\hline
Mini-flux rope  & &  & \\
\hline
LE [UT]        & 4/19 07:05 &  4/20 05:55 &  4/20 07:17 \\
TE [UT]        & 4/19 08:59 &  4/20 06:55  & 4/20 08:05  \\
$\Delta T_{\mathrm{SFR}}$ [h] & 1.9    &  0.82        & 0.68   \\ 
$\langle \mathrm{B}_{\mathrm{SFR}} \rangle$ [nT] & $13.3\pm1.0$    &  $8.8\pm0.35$ & $10.2\pm0.61$  \\
$\theta_{\mathrm{SFR}}$ & $0^{\circ}$ & $21^{\circ}$ & $13^{\circ}$  \\
$\phi_{\mathrm{SFR}}$ & $75^{\circ}$ & $118^{\circ}$ & $79^{\circ}$  \\
$p$         & 0.0017 & 0.078 & 0.17  \\
\hline
\end{tabular}
\label{tab:times}
\end{table*}

{At Wind, the shock leading  the ICME is identified as a simultaneous jump of the plasma parameters and magnetic field magnitude, while at BepiColombo and Solar Orbiter from the sharp increase in the magnetic field magnitude only. At each location the magnetic field magnitude increases from $\sim 3$ nT to about 6--7 nT. The Wind observations show that the solar wind speed jumps from about 280 km/s to 340 km/s and the density from $\sim 5$ cm$^{-3}$ to $10$ cm$^{-3}$.}

{Key shock parameters are described below \citep[for methods see, e.g.,][]{Kilpua2015}. In Table \ref{tab:times}, the direction of the shock normal ($\mathbf{n}_{\mathrm{sh}}$) at each spacecraft computed using the magnetic coplanarity method is provided. Also listed are the shock angles, i.e., the angle between the shock normal and the upstream magnetic field direction ($\theta_{\mathrm{Bn}}$), and the downstream-to-upstream magnetic field ratio ($B_{\mathrm{d}}/B_{\mathrm{u}}$). When calculating the upstream and downstream values, and interval consisting of two minutes before and after the shock was excluded to avoid the shock transition layer. The durations of the intervals were chosen such as to be long enough to average out any fluctuations related to waves and turbulence, but short enough to avoid disturbances not associated with the shock. The analysis indicates that the shock was quasi-perpendicular at Wind, and at the limit of quasi-parallel and quasi-perpendicular shock ($\theta_{\mathrm{Bn}}=45^{\circ}$) at Solar Orbiter and BepiColombo.
For Wind we calculated also the shock speed, $v_\mathrm{sh} =353$ km/s. The obtained upstream Alfv\'{e}n speed is 30.6 km/s and the upstream sound speed is 25.3 km/s, yielding the upstream magnetosonic speed $v_\mathrm{up,ms}$ = 40.0 km/s. This gives the shock magnetosonic Mach number $M_\mathrm{ms}=1.9$. The shock had thus a slow-speed and was weak. }

{At all three spacecraft, the shock was followed by a period of larger-amplitude fluctuations in the magnetic field direction; at Wind, it can be seen that the proton density was significantly enhanced (up to $\sim 45$ cm$^{-3}$). These are typical signatures of a compressive ICME sheath region \citep[e.g.,][]{Kilpua2017}. The subsequently observed ejecta featured a smoothly rotating magnetic field direction over a large angle, low field variability, and enhanced magnetic field magnitudes ($\sim 23$ nT at Solar Orbiter and $\sim 16$ nT at Wind and BepiColombo), indicating a flux rope configuration. These signatures are consistent with the presence of a magnetic cloud  \citep[e.g.,][]{Burlaga1981,Burlaga1988}. Counterstreaming electrons were not observed; although they are a standard magnetic cloud signature, a significant numbers of magnetic clouds do not display them, or they are only present within a subsection of the structure \citep[e.g.,][]{Shodhan2000}. The earlier candidate leading edge time (LE1, solid red line) coincided with a sharp change in the magnetic field direction, with a prominent change in the $B_N$ component from positive to strongly negative. This abrupt directional change coincided with a considerable increase in the magnetic field magnitude and (as seen at Wind) a sharp and deep drop in density. The later possible leading edge time (LE2, dashed red line) marks the time when the more coherent field rotation started. Between LE1 and LE2, the magnetic field components were smoother than between the shock and LE1, but sharp changes in $B_T$ occurred. The $R$-component of solar wind velocity ($V_R$) was approximately steady, while from LE2 onwards, it declined towards the trailing part of the cloud. $V_T$ and $V_N$ show deflections of up to $50$ km/s, which can indicate deflection of the solar wind around the magnetic cloud obstacle.  Based on the analysis above, with further supporting evidence given in Section 3.3, we consider that the interval between LE1 and LE2 belonged to the magnetic cloud, but had been distorted during interplanetary propagation \citep[see e.g.,][]{Kilpua2013}.

{The earlier trailing boundary time (TE1, solid blue line) of the magnetic cloud in Figure \ref{fig:FIG_Overview} is placed at the time when the relatively smooth field rotation ended and (at Wind) when the density increased. After TE1, there were large-amplitude field variations, but signatures of coherent rotation in field components continued until the second trailing boundary (TE2, dashed blue line), which marked the arrival of a faster solar wind stream at Wind. The interval between TE1 and TE2 represents a region of plasma and magnetic field that was disturbed by the trailing solar wind. Since our focus is on the sheath of this ICME, we do not discuss the end part of the cloud further here.}

\subsection{Energetic ion observations}

{In this section, energetic ion observations at the different spacecraft are described. Data are available from the ACE spacecraft at L1 and from Solar Orbiter. Energetic electron observations were available from all three locations (L1, BepiColombo and Solar Orbiter), but none of them showed significant enhancement. We also note that no flares (nor type III radio bursts) were observed during the event that could have affected the observations.}

\begin{figure*}[ht]
\centering
\includegraphics[width=0.9\linewidth]{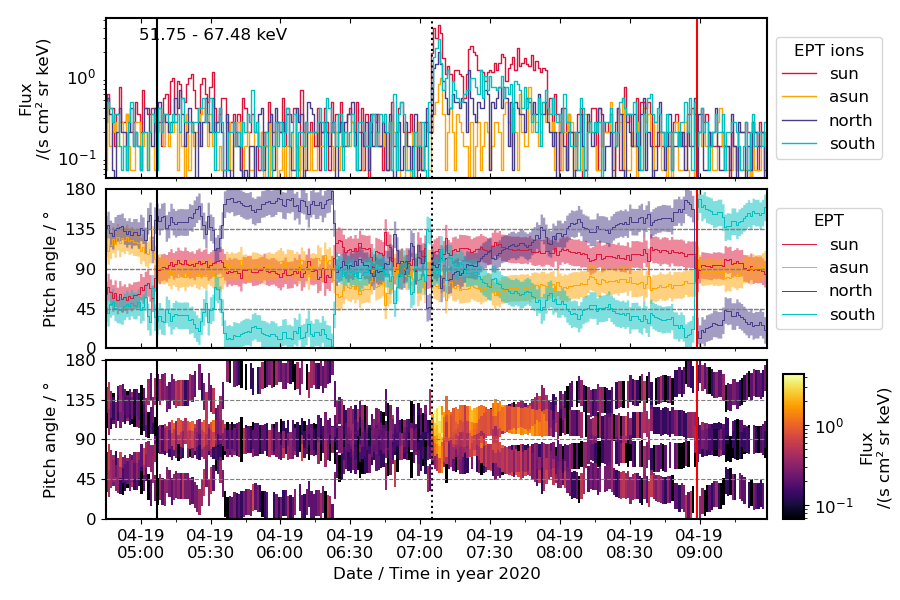}
\caption{Energetic ion observations by SolO/EPT within the mini flux rope (see discussion in Section 3.3.3). From top to bottom: 59 keV ion intensities observed in the four viewing directions of EPT, corresponding pitch-angle ranges covered by these viewing directions, and pitch-angle distribution of the 59 keV ions (intensity in color coding). The three vertical lines mark the times of the shock (black solid line), the beginning of the mini flux rope (dotted line), and the first leading edge (LE1, marking the end of the mini flux rope, red solid line), respectively.}
\label{fig:FIG_EPT_PAD_zoom}
\end{figure*}

Fig. \ref{fig:FIG_EPT_PAD_zoom} presents energetic ion observations of the Solar Orbiter EPT instrument. The top panel shows the sectored 59 keV ion intensities and the middle panel illustrates the corresponding pitch angle ranges covered by EPT's four viewing directions. Information of the two upper panels is combined in the bottom panel showing the color-coded pitch angle distribution of the ions. From left to right the vertical lines mark the time of the shock and the beginning and end of a distinct structure henceforth described as the mini flux rope which we will discuss in detail in Section 3.3.3.

While only a barely noticeable energetic particle event is observed at the time of the shock an increase in energetic ion fluxes is observed by Solar Orbiter coincident with entry into the mini flux rope lasting for about one hour. The increase reaches a maximum energy of $\sim$100~keV. Unfortunately, during the time of this increase the pitch-angle coverage is not ideal due to the non-Parker magnetic field orientation so that only pitch angles close to $90^{\circ}$, i.e. perpendicular to the magnetic field, are covered. It is therefore difficult to determine whether a significant anisotropy was present. The small intensity differences between the different viewing directions, however, suggests that some anisotropy could be present, i.e., increase of flux at pitch-angles $\sim 90^{\circ}$.  However, low energy ions are subject to the Compton-Getting effect \citep{Ipavich1974}, which causes an increase (decrease) of intensities in the sunward (anti-sunward) looking direction which can mimic an anisotropy in the non-corrected data. 

We applied a correction for this Compton-Getting effect (i.e., a transformation to a frame moving with the solar wind velocity). The approach is described in more detail in Appendix~\ref{app:b}. Using this transformation, we obtained the differential energy spectra of energetic ions for a frame moving with the solar wind,
based on the spacecraft frame measurements by SolO/EPT for the four viewing directions of EPT (see Fig.~\ref{fig:FIG_EPT_spec}). 
Because the plasma instrument onboard Solar Orbiter was not operating during the observation period, we used the shifted observations from Wind spacecraft. Figure~\ref{fig:FIG_Overview_WIND} indicates that the solar wind velocity is mainly radial and rather constant with a value of 350 km/s between the shock (black vertical line) and the leading edge (solid red line). Figure~\ref{fig:FIG_EPT_vspace} shows the intensities in the solar wind frame in the $(v_\parallel',v_\perp')$ plane. The observations cover well only the pitch-angle region around 90$^\circ$, and there are no significant signs of anisotropy in this region of velocity space. Note also, that the intensity and pitch-angle coverage are temporally variable during the hour depicted, and that the intensity uncertainties increase significantly above proton speeds of 4000 km/s. The observations are consistent with the observed anisotropies in the spacecraft frame to be the result of the Compton--Getting effect. 

\begin{figure*}[ht]
\centering
\includegraphics[width=0.9\linewidth, viewport=5 7 700 329, clip]{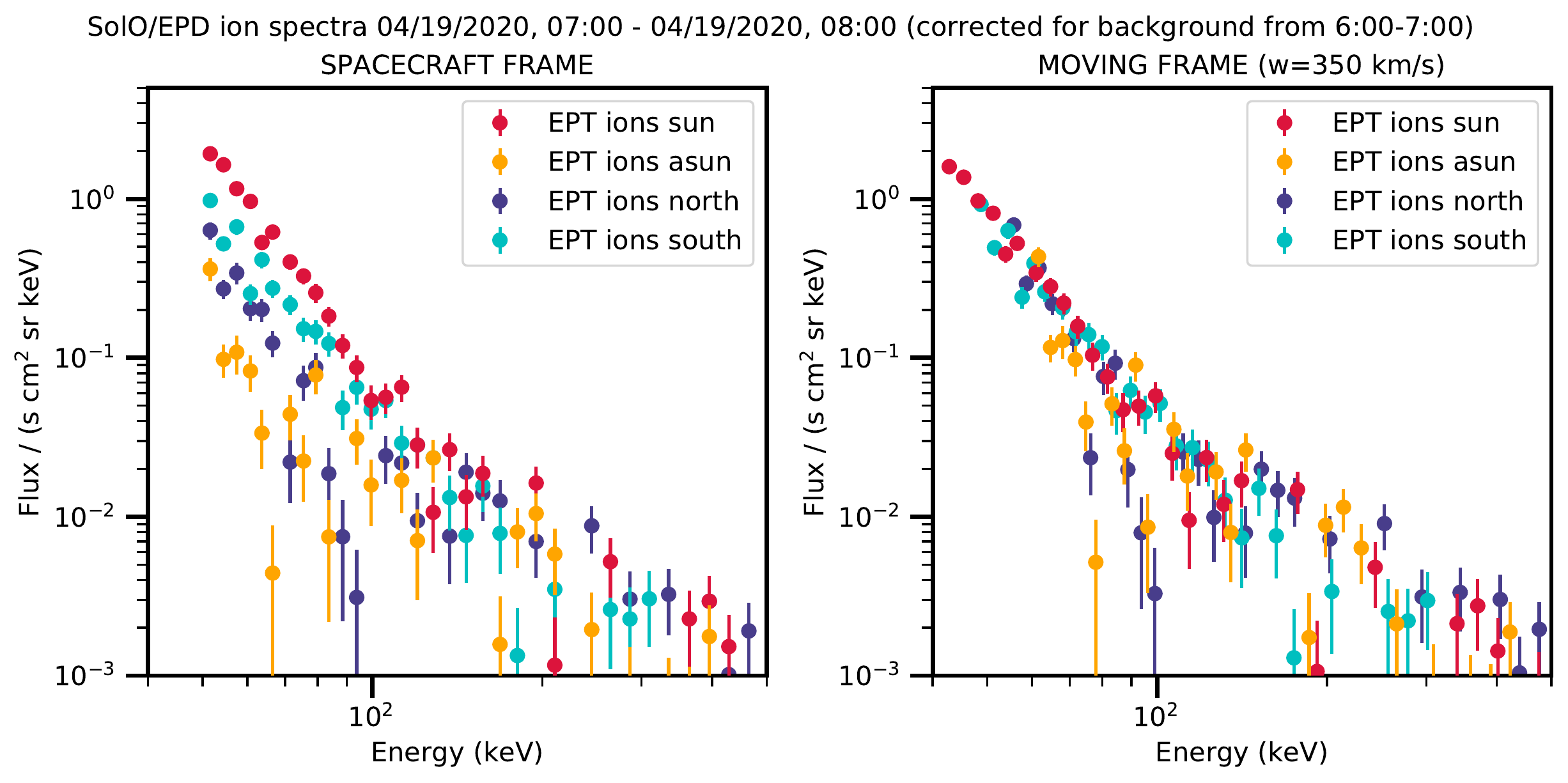}
\caption{Differential energy spectra of energetic ion observations by SolO/EPT for the four viewing directions of EPT. The time interval is 7:00 to 8:00~UT on April 19, 2020, covering the increase of energetic ion fluxes within the mini flux rope (see Fig.~\ref{fig:FIG_EPT_PAD_zoom} and discussion in Section 3.3.3). Left: Measured in the spacecraft frame. Right: Transformed to a frame moving with the solar wind for a radial speed $w=350$ km/s.
}
\label{fig:FIG_EPT_spec}
\end{figure*}

\begin{figure}[ht]
\centering
\includegraphics[width=0.99\linewidth, viewport=5 45 480 260, clip]{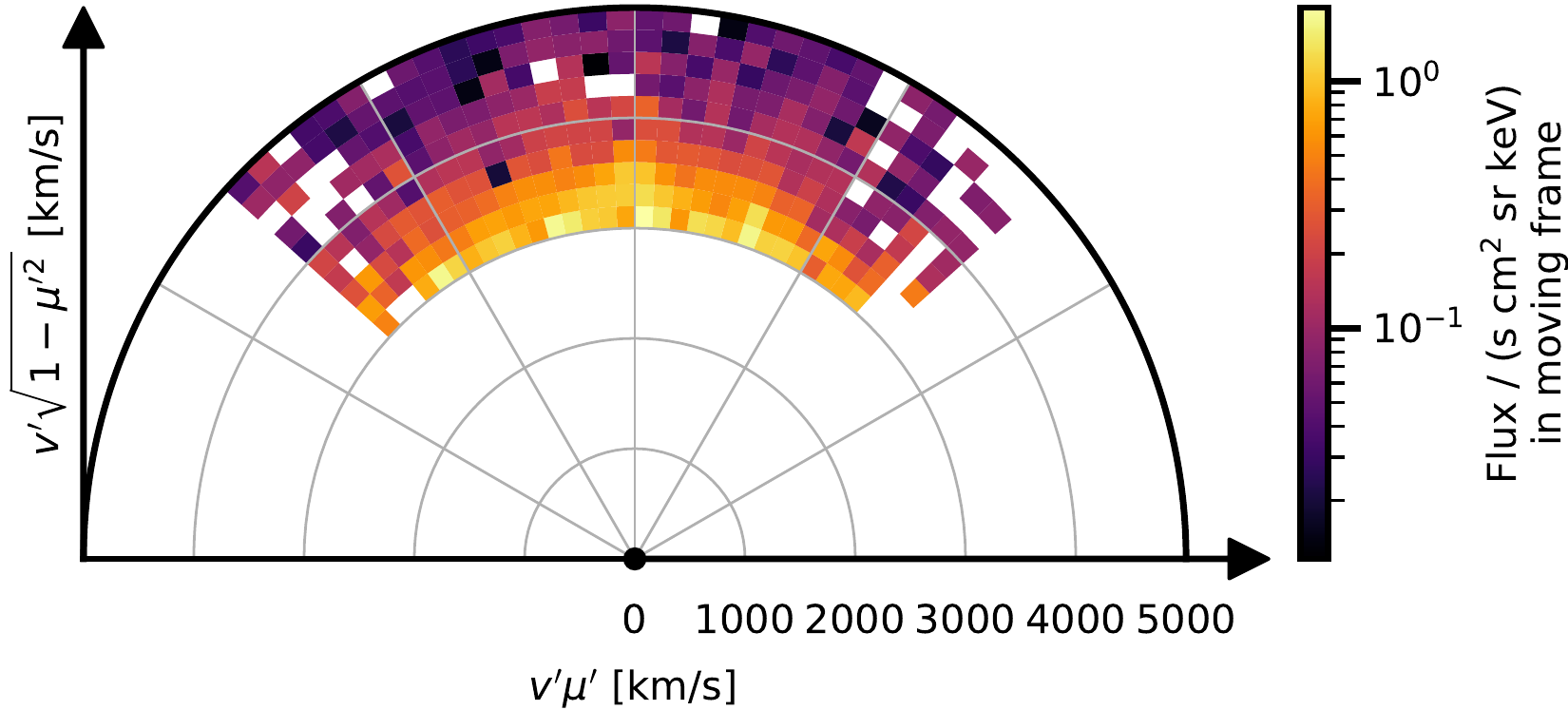}
\caption{SolO/EPT ion observations in a frame moving with the solar wind for a radial speed $w=350$ km/s, at proton speeds between 3000 and 5000 km/s, respectively. Here $v'$ and $\mu'$ are the ion speed and pitch-angle cosine in the moving frame of reference, respectively. 
The time interval is 7:00 to 8:00~UT on April 19, 2020, covering the increase of energetic ion fluxes within the mini flux rope (see Figs.~\ref{fig:FIG_EPT_PAD_zoom} \& \ref{fig:FIG_EPT_spec} and discussion in Section 3.3.3).}
\label{fig:FIG_EPT_vspace}
\end{figure}

Figure \ref{fig:FIG_ACE_protons} shows energetic ions as detected by the ACE/EPAM instrument's Low Energy Magnetic Spectrometer (LEMS) for four energy channels covering energies from 47 to 310 keV. There is a clear enhancement of energetic ions in the sheath observed on April 20, 2020 between 01:55 UT to 03:20 UT clearly visible for all shown channels except the highest channel, i.e. up to 190 keV. The magnetic field components shown in the figure show that ACE detected a very similar field behaviour as Wind (Figure \ref{fig:FIG_Overview}) as expected since both spacecraft were at L1. Sharp magnetic field turnings are observed bounding the enhancement, these are discussed in more detail in Section 3.3.2.

\begin{figure}[ht]
\centering
\includegraphics[width=0.85\linewidth]{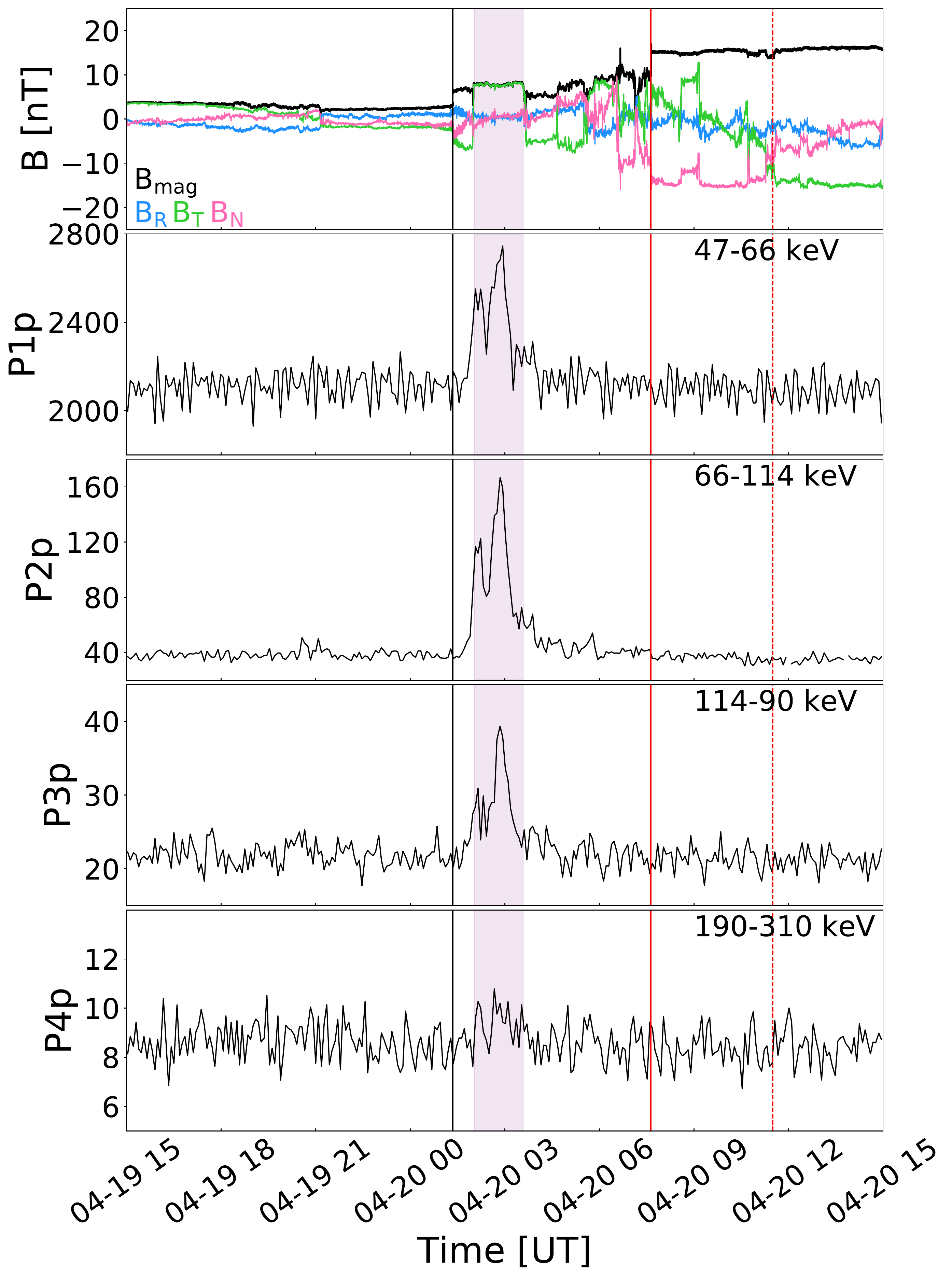}
\caption{The magnetic field and energetic ions at ACE (located at L1). The top panel shows the magnetic field magnitude (black) and three magnetic field components in RTN (blue: $B_R$, green: $B_T$, pink: $B_N$ . The following panels give the energetic ions observations for four energy ranges (P1p -P4p with energies shown in each panel). The units are $\textrm{cm}^{-2}\,\textrm{sr}^{-1}\,\textrm{s}^{-1}\,\textrm{MeV}^{-1}$. The shock is shown by the vertical black line and the magnetic cloud leading edge by the red solid line.}
\label{fig:FIG_ACE_protons}
\end{figure}

\subsection{Sheath fluctuations and small-scale structures}

{In this section, smaller-scale structures and magnetic field fluctuations in the sheath regions at the three locations are described in more detail. We focus on periods coinciding with the energetic ion enhancements discussed in Section 3.2 at Solar Orbiter and L1. These ion enhancements occurred at different locations in the sheath: just adjacent to the magnetic cloud leading edge (LE1) at Solar Orbiter, while closer to the shock at Wind. The corresponding sheath substructures at BepiColombo are also investigated, although energetic ion data were not available at the spacecraft.}

\subsubsection{Heliospheric current sheet crossings} 

{As discussed in Section 3.2, the period of energetic ions enhancement at ACE was bounded by two sharp changes in the magnetic field direction. We now explore the features of this field structure at L1, and also the features of similar structures at BepiColombo and Solar Orbiter.}

{Figure \ref{fig:WIND_angles} shows the zoom-in for Wind with the magnetic field, plasma and suprathermal  electron pitch angle distribution data and Figure \ref{fig:FIG_Fluc_Angles} shows the zoom-in for the three spacecraft investigated with the magnetic field data. The dash-dotted lines in panels showing the RTN longitude ($\phi_B$) correspond to the nominal Parker spiral angles at 0.8 AU (Solar Orbiter) and $\sim 1$ AU (L1 and BepiColombo. At both locations, a solar wind speed of 325 km/s (the average during the sheath at Wind) was used to calculate the angles. The Parker spiral angles were estimated to be $45^{\circ}$ at 0.8 AU and $51^{\circ}$ at 1 AU; in terms of $\phi_B$, the towards/away spiral angles with respect to the Sun were thus $135^{\circ}$/$315^{\circ}$ at 0.8 AU and $129^{\circ}$/$309^{\circ}$ at 1 AU. The solid horizontal lines in the $\phi_B$ panels indicate the sector boundaries, at $\pm 90^{\circ}$ of the spiral angles: the IMF was thus in the towards sector for $45^{\circ}<\phi_B<225^{\circ}$ at 0.8 AU ($51^{\circ}<\phi_B<231^{\circ}$ at 1 AU), and otherwise in the away sector.} 

\begin{figure}[ht]
\centering
\includegraphics[width=1\linewidth]{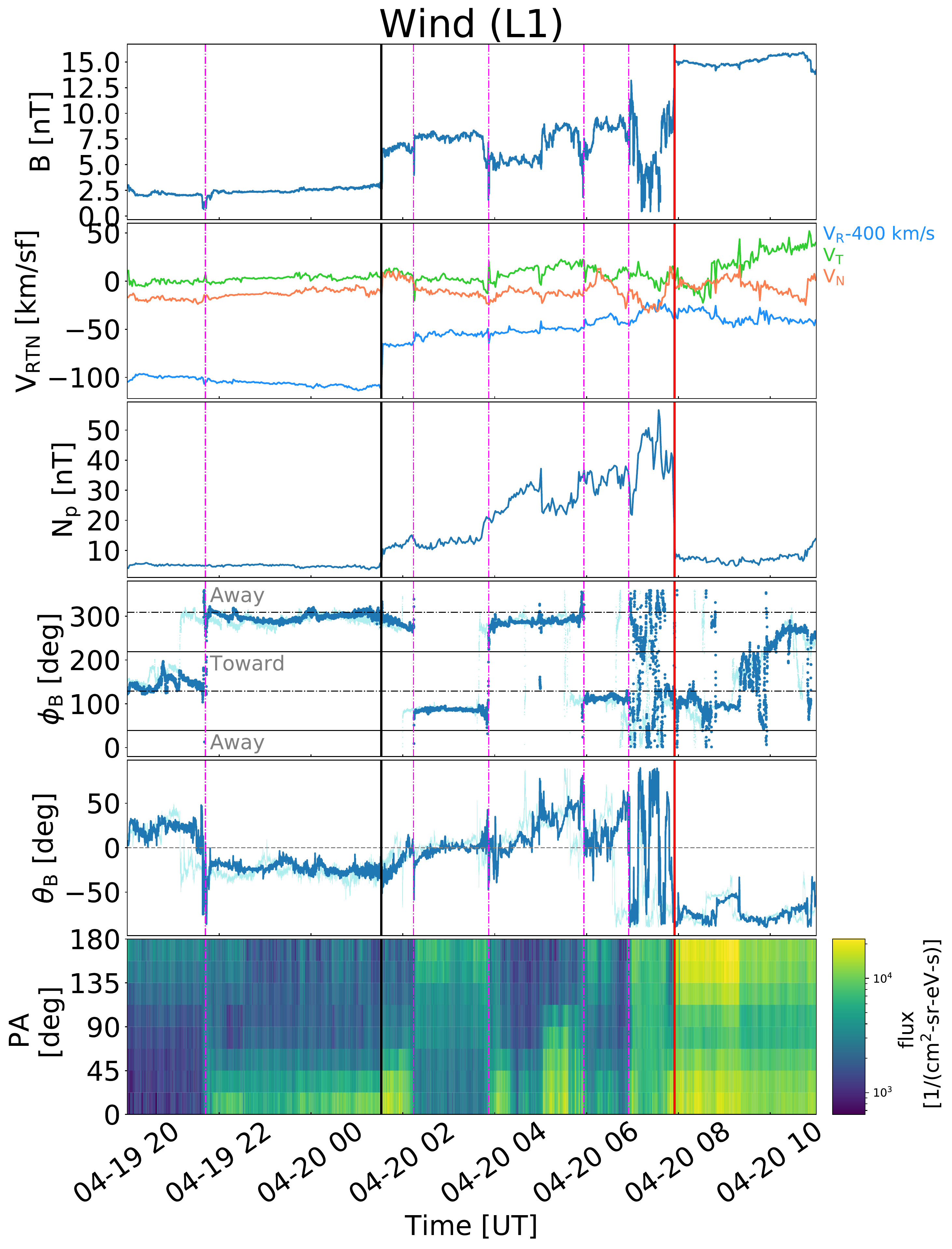}
\caption{A zoom-in to the sheath at L1. The panels show, from top to bottom: The magnetic field magnitude; solar wind velocity components in RTN coordinates; density; magnetic field azimuth and latitude angles (dark blue: Wind, light blue: ACE; the ACE data is not time-shifted); and the pitch angle distribution of 265 eV electrons. 
Vertical black and red lines indicate the shock and leading edge (LE1) times, respectively, and magenta lines indicate relatively sharp field changes of interest. Nominal away and towards sectors of the IMF are indicated in the $\phi_B$ panels, i.e. towards sector is between the solid horizontal lines (see text for details).
 HCS crossings are shown by pink dash-dotted lines. In the azimuth angle plot, the two dashed-dotted horizontal lines indicate the Parker Spiral direction and  solid horizontal lines indicate the sector boundaries (see text for details).}
\label{fig:WIND_angles}
\end{figure}

\begin{figure*}[ht]
\centering
\includegraphics[width=1\linewidth]{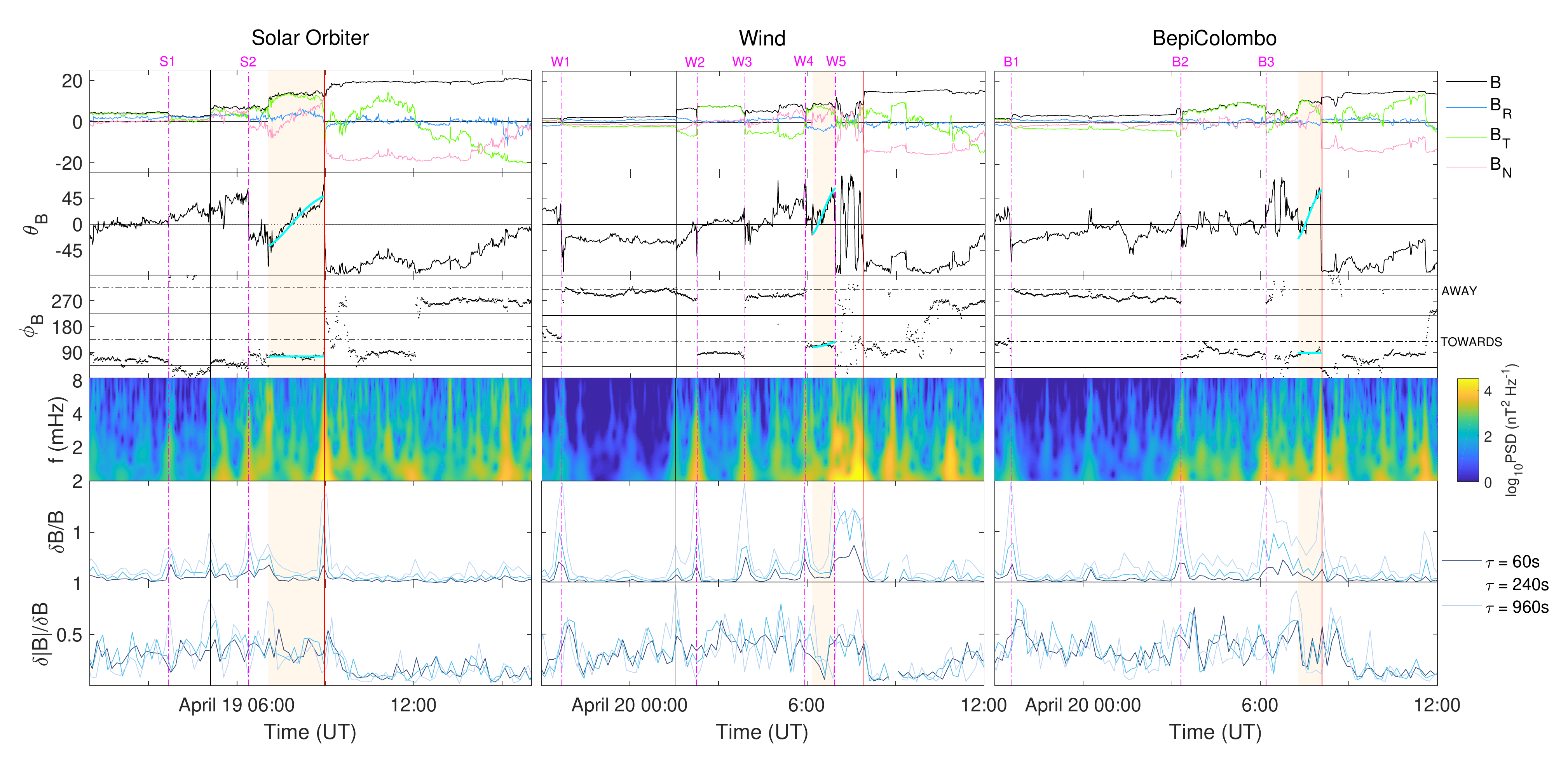}
\caption{The sheath at Solar Orbiter, Wind and BepiColombo. From top to bottom, the panels show: the B-field magnitude and components in RTN coordinates; RTN B-field azimuth angle; RTN B-field latitude angle; wavelet PSD of B-field fluctuations at 1--8~mHz; normalised B-field fluctuation amplitude at three timescales; and fluctuation compressbility. The vertical lines as same as in Figure \ref{fig:WIND_angles}, the shock is shown by the solid black line and the magnetic cloud leading edge time (LE1) by the solid red line, and the magenta lines indicate relatively sharp field changes of interest.  Orange shading indicates the location of a small-scale flux rope identified within the sheath at each spacecraft, and cyan lines show Gold-Hoyle flux rope fits to these intervals.}
\label{fig:FIG_Fluc_Angles}
\end{figure*}

{Vertical magenta dash-dotted lines in Figure \ref{fig:FIG_Fluc_Angles} and Figure \ref{fig:WIND_angles} mark the possible HCS crossings \citep[e.g.,][]{Smith2001}. They are characterised by a sharp change in the field direction consistent with a change in IMF sector (i.e., a change in $\phi_B$ of $\sim 180^{\circ}$), and, additionally at Wind, by the simultaneous change in suprathermal electron pitch angle distribution from $0^{\circ}$ to $180^{\circ}$, or vice versa.}

{An HCS crossing occurred upstream of the shock at Wind (W1) and BepiColombo (B1), at which the IMF changed from the towards to away sector, and at which suprathermal electrons at Wind changed from an anti-parallel to parallel flow relative to the magnetic field direction. At Solar Orbiter, the upstream field was in the towards sector very close to the nominal sector boundary, then drifted into the away sector at S1 ($\phi_B$ changing by $\sim 45^{\circ}$), just before the shock.}

{Downstream of the shock at Solar Orbiter, the IMF was again close to a sector boundary, before shifting into the towards sector at S2. As at S1, the change in $\phi_B$ at S2 was relatively small and not consistent with a clear HCS crossing. At L1, in contrast, several clear HCS crossings occurred within the sheath. HCS crossings at W2 and W3 exhibited signatures of magnetic reconnection exhausts \citep[e.g.,][]{Gosling2006,Lavraud2009,Lavraud2020}. Figures presented in the appendix show the signatures of this reconnection, including velocity jets (in $V_T$) and dips in the magnetic field magnitude. In addition, the appendix figures show that W2 was disconnected (i.e., there was a strahl dropout), while W3 displayed bidirectional suprathermal electrons. The subsequent HCS at W4 was also reconnecting and featured a heat flux drop-out, while the HCS at W5 was not reconnecting, and was associated with counterstreaming electrons. Between W5 and the magnetic cloud leading edge LE1 (red line), there were several large-amplitude, out-of-ecliptic magnetic field fluctuations and sector changes. At BepiColombo, two clear HCS crossings are marked at B2 and B3 within the sheath. After B3, the magnetic field direction fluctuated between the two sectors for a short interval, and then remained in the towards sector up to LE1.} }

\begin{figure*}[ht]
\centering
\includegraphics[width=1\linewidth]{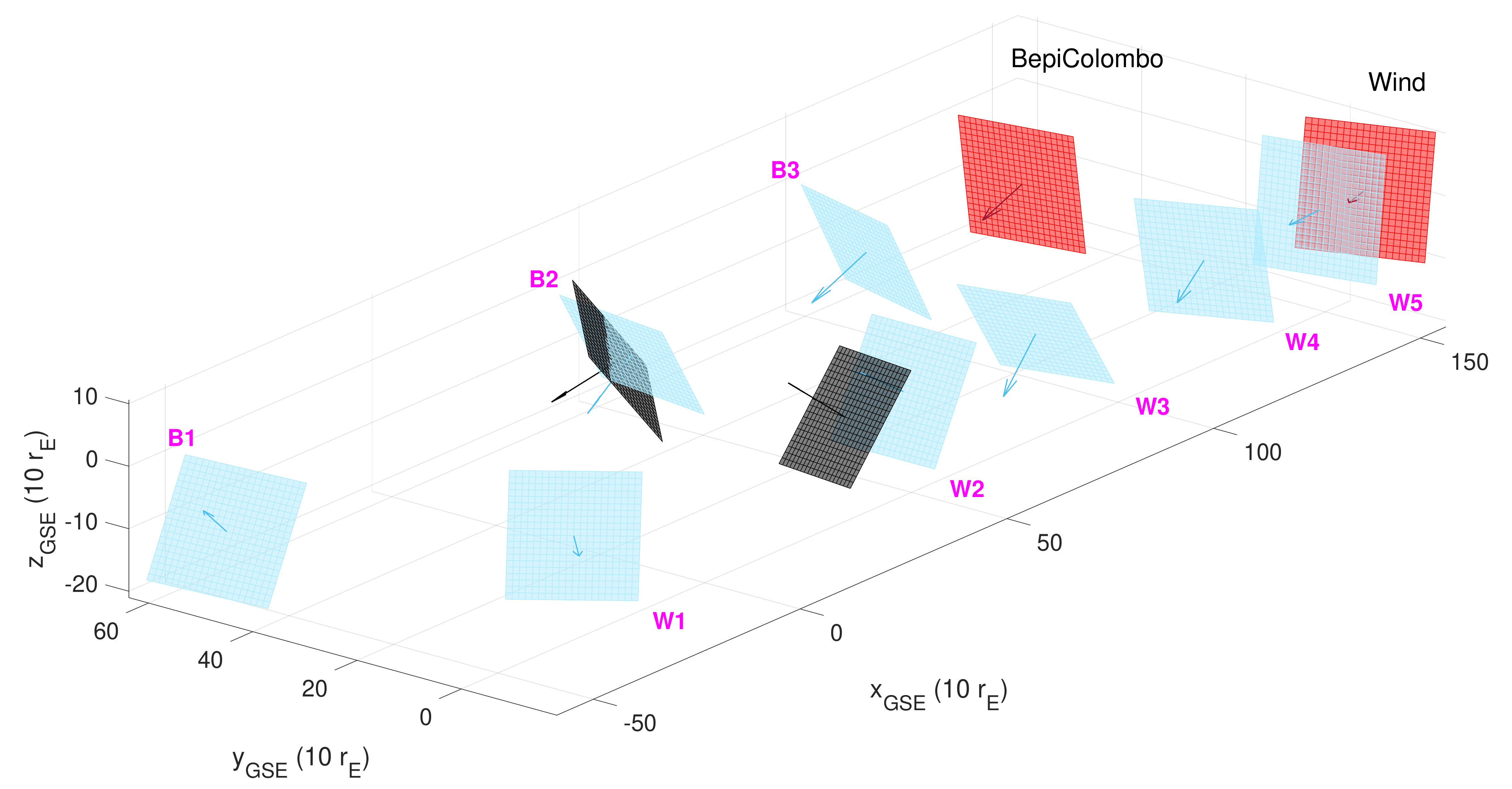}
\caption{Schematic of the current sheet (blue), shock (black), and magnetic cloud leading edge LE1 (red) planes at BepiColombo and L1.}
\label{fig:3Dsheath}
\end{figure*}

{Based on visual inspection of the data, we suggest that W1--W3 at Wind corresponded to B1--B3 at BepiColombo, respectively. W4 and W5 had no clear counterparts at BepiColombo, although they may have corresponded to the interval containing multiple sector crossings immediately after B3. The local orientations of the HCS, shock and ICME leading edge (LE1) surfaces at BepiColombo and Wind are represented by square planes in Figure \ref{fig:3Dsheath}, with arrows showing the surface-normal vectors. The HCS and ICME leading edge normals were estimated with minimum variance analysis applied over 20-min intervals centred on the feature in question (a 10-min interval for B2 to avoid the nearby shock). The shock normals were obtained via the methods described in section 3.1. Figure \ref{fig:3Dsheath} shows the estimated spatial locations of the various structures at the shock arrival time at Wind. While the orientations of some structures were broadly consistent at the two L1 spacecraft (e.g., B3 and W3), in the sense that they would be consistent with some global, 3-dimensional structure that varied only gradually over the (relatively small) spacecraft separation distance, it can be seen that other structures (e.g., the shock) showed much greater variation over the same distance scale.}

{We note finally that the zoom-in plots in the appendix show that the $V_T$ jets are oppositely directed at W2 and W3. Assuming that the L1 spacecraft crossed the exhaust from the opposite sides of the same X-line (as would be plausible based on their close proximity and similarity), these opposite directions imply that the HCS was folded or 'wavy' \citep[e.g. see Figure 7 in][]{Mistry2015}. The field changes for W2 and W3 are also step-wise, indicating that the current sheet crossed by the spacecraft was bifurcated \citep[e.g.,][]{Mistry2015b}.}

\subsubsection{Sheath fluctuations}

{The fourth panel of Figure \ref{fig:FIG_Fluc_Angles} shows the wavelet power spectral density (PSD) of magnetic field fluctuations in the 1--8~mHz frequency range, with the local mean field subtracted. The bottom two panels give the time series of normalised magnetic field fluctuation amplitudes, $|\delta \mathbf{B}|/\delta B$, and the magnetic field fluctuation compressibility, $\delta |\mathbf{B}|/\delta B$. Both quantities are calculated over successive (non-overlapping) 10-min intervals. The magnetic field fluctuation amplitudes are defined as $|\delta \mathbf{B}| = |\mathbf{B}(t) - \mathbf{B}(t+\tau)|$, where $\tau$ is the time lag between two sample points, i.e., the fluctuation timescale. The fluctuations are calculated for $\tau=$ 60, 240 and 960~s, all of which correspond to magnetohydrodynamic inertial scales in terms of turbulence phenomenology i.e. at scales larger than the proton gyroscale. The wavelet PSD frequencies in the fourth panel also fall within the inertial range.}

{Across the frequency range investigated and at all three spacecraft, there was a broadband enhancement in fluctuation PSD in the sheath compared to the upstream solar wind. In contrast, and excluding localised spikes, there was little overall increase in $|\delta \mathbf{B}|/\delta B$ from upstream to downstream. While the increase in PSD indicates that fluctuation amplitudes were larger in the sheath, the relatively flat $|\delta \mathbf{B}|/\delta B$ suggests that the fluctuation amplitudes were largely proportional to the background mean field, $B$, and that the increased fluctuation amplitudes in the sheath were caused by an increase in $B$. This indicates that the level of turbulence at the sampled frequencies was not particularly enhanced in this sheath when compared to the solar wind ahead. Fluctuation PSD was somewhat lower in the upstream solar wind at BepiColombo and Wind than at Solar Orbiter, consistent with fluctuation amplitudes falling with heliocentric distance.}

{The figure also clearly shows sharp and local enhancements in PSD and $|\delta \mathbf{B}|/\delta B$ coinciding with HCS or sector crossings (vertical pink dash-dotted lines). The strongest enhancement away from the HCS crossings occurred just before the leading edge of the magnetic cloud at Wind and BepiColombo, and particularly so at Wind, where large-amplitude, out-of-the-ecliptic field fluctuations were present. These near-leading edge fluctuations were most likely associated with field line draping around the magnetic cloud \citep[e.g.,][]{Gosling1987b,Comas1988}.}

{Fluctuation compressibility was at approximately the same level in the upstream solar wind and the sheath at all locations, and was independent of the three timescales sampled. Upon entering the magnetic cloud (at LE1; red solid vertical line in Figure \ref{fig:FIG_Fluc_Angles}), compressibility dropped significantly. The lower compressibility of fluctuations in magnetic clouds versus sheaths (or upstream solar wind) was previously reported by \cite{Moissard2019} and \citet{Kilpua2021}. This drop in compressibility supports the conclusion drawn in Section 3.1, namely that the region between LE1 and LE2 is part of the ejecta, albeit with disturbed field and plasma properties.}

\subsubsection{Mini flux rope}

{A distinctive feature is visible at Solar Orbiter at the exact time when the energetic ion enhancement was observed. The magnetic field shows a structure with smoothly rotating field components, as indicated by the three top left panels of Figure \ref{fig:FIG_Fluc_Angles} within the orange-shaded region. The $B_N$ rotation from negative to positive is particularly prominent. The strongest ion enhancement coincide at the front part of this structure when the field rotates from negative to zero.}

{This structure resembles a small-scale magnetic flux rope. These are regularly observed in the slow solar wind \citep[e.g.,][]{Moldwin2000,Yu2014} and have also been previously reported in ICME sheaths \citep[e.g.,][]{Kilpua2020}. This mini flux rope (mini-FR) starts on April 19, 07:05 UT and extends until the sharp change in the field direction marking LE1, approximately 2 hours later. The leading and trailing edge times of the mini-FR and some of its key parameters are collected in Table \ref{tab:times}.} 

{The fit of the Gold-Hoyle flux rope model \citep{Gold1960,Farrugia1999} to this interval is shown by the cyan lines overlaying the field angle panels in Figure \ref{fig:FIG_Fluc_Angles}. The model solution, which assumes a constant twist through the flux rope, produced a very good match with the observed magnetic field profiles. The fitting yields a flux rope with a positive magnetic helicity sign and very low axis inclination with respect to the ecliptic plane, with an axis longitude and latitude of $(\theta_{\mathrm{FR}},\phi_{\mathrm{FR}})=(\sim 0^{\circ},75^{\circ})$. The impact parameter (i.e., the ratio of the closest approach distance of the spacecraft from the flux rope axis to the flux rope radius) of $p=0.0017$ suggests that Solar Orbiter crossed this mini-FR very close to the axis.}

{Signatures of the mini-FR are also present in the BepiColombo and L1 spacecraft data. Figure \ref{fig:FIG_Fluc_Angles} shows that, adjacent to LE1 at both spacecraft (orange-shaded intervals), there were periods when the magnetic field components showed comparable variations to the mini-FR at Solar Orbiter. The leading part of the mini-FR appears comparatively distorted or eroded, with $B_N$ rotating from zero to positive field values, instead of the negative to positive rotation detected at Solar Orbiter. I.e., the part at Solar Orbiter that was associated with the ion enhancement was distorted at L1 and BepiColombo. In addition, at L1, the mini-FR ends about 1 hour before the magnetic cloud starts. The interval in between is marked by large amplitude field variations and decreased magnetic field magnitude. The durations of these mini-FR intervals at BepiColombo (41 minutes) and L1 (49 minutes) are shorter than at Solar Orbiter. However, it does not appear that the mini-FR was significantly further compressed at the leading edge since the average magnetic field magnitudes at BepiColombo (10.2 nT) and L1 (8.8 nT) were lower than at Solar Orbiter (13.3 nT). Gold-Hoyle fits to the FR intervals at Wind and BepiColombo give axis orientations of $(\theta_{\mathrm{FR}},\phi_{\mathrm{FR}})=(21^{\circ},118^{\circ})$ and $(\theta_{\mathrm{FR}},\phi_{\mathrm{FR}})=(13^{\circ},79^{\circ})$, respectively, and impact parameters of 0.078 and 0.17, respectively. The mini-FR thus also had low inclinations at L1 and at BepiColombo, but was intersected a bit further from the axis by the spacecraft. Furthermore, Figure \ref{fig:FIG_Fluc_Angles} shows that the mini-FR featured decreased levels of magnetic field fluctuations at all spacecraft.}

{The most probable cause of the differences in the mini-FR properties at the different locations is radial evolution from 0.8 to 1 AU, and the relatively large separation of the spacecraft along the HEEQ $Y$ direction (916 $R_E$ or 0.039 AU for Solar Orbiter and L1, and 1508 $R_E$ or 0.064 AU for Solar Orbiter and BepiColombo; see Table \ref{tab:location}). Similarly to the HCS structure described previously at Wind and BepiColombo, the mini-FR properties (e.g., axis orientation) appeared to change over length scales smaller than the spacecraft separation. Using the average speed of the mini-FR of 355 km/s at Wind gives a radial width for this structure of 401 $R_E$ (0.017 AU) at Solar Orbiter, smaller than the separation between Solar Orbiter and L1/BepiColombo along the HEEQ Y direction. Flux ropes in the solar wind typically have considerably larger lateral than radial dimensions. }


\section{Discussion} \label{sec:discussion}

{We have analysed the internal, small-scale structure of an ICME-driven sheath region and the relation of this substructure to energetic ion enhancements. The fast forward shock preceding the sheath was weak (magnetosonic Mach number 1.9) and had a slow propagation speed (353 km/s in the spacecraft frame). Energetic particle enhancements were not coincident with the shock, but occurred later in the sheath.  No drastic difference in the level of MHD-scale turbulence in terms of normalised magnetic field fluctuation amplitudes was found between the upstream solar wind and the sheath. In addition, the level of compressibility in magnetic field fluctuations was similar between the upstream wind and sheath.}

{Distinct structures in the sheath at both Solar Orbiter and L1 were coincident with the ion enhancements. At L1, the ion enhancement occurred relatively close to the shock and between two reconnecting current sheets associated with the HCS. Their characteristics imply a likely mesoscale (several hundreds $R_E$) fold or wave in the HCS that had recently been swept across the shock and compressed into the sheath. No similar structure was identified at Solar Orbiter, but similar mesoscale features occurred at BepiColombo, although with a somewhat greater spatial extension and with different orientations. Significant evolution in HCS structure has been identified previously in the inner heliosphere, most recently by \cite{Szabo2020} in their comparison of PSP and L1 observations. Such structures could imply that the shock has recently propagated over field lines that have complicated geometries, which can facilitate particle acceleration if the shock crosses a single field line in multiple locations \citep{Sandroos2006}. Thus, for the ion enhancement at L1 a relation to the shock cannot be excluded.}

{At Solar Orbiter, the ion enhancement was observed inside a well-defined mini-flux rope (FR). This mini-FR was also present at both L1 and BepiColombo, although the structure was less coherent and of shorter duration at these spacecraft. In particular its front part that was associated with the ion enhancement at Solar Orbiter was distorted at L1/BepiColombo. The Gold-Hoyle uniform-twist fitting yielded broadly consistent results for the mini-FR axis orientation at the different locations. The impact parameter from the fitting indicates that Solar Orbiter crossed the mini-FR closer to the central axis than the other spacecraft.}

{Previous studies have suggested that coherent magnetic field structures in the solar wind can modulate energetic particle fluxes. For example, \cite{Neugebauer2015} connected sharp changes in the energetic particle fluxes to prominent changes in the magnetic field and plasma in the solar wind, which they identified as tangential discontinuities or flux tube boundaries. Recent simulation studies have also shown that the passage of current sheets past shock waves can make the current sheets unstable and lead to particle acceleration \citep[e.g.,][]{Nakanotani2020J}. In contrast, the ion enhancement observed here was confined between two reconnecting HCS crossings.}

{Acceleration associated with merging or contraction of flux tubes or ropes near CMEs, the HCS and slow-fast stream interaction regions (SIRs) has been suggested in previous studies as a possible mechanism responsible for producing energetic ions \citep[e.g.,][]{Khabarova2016,Tessein2016,Roux2018}. The location of the mini-FR close to the leading edge of the magnetic cloud suggests that it was swept there relatively early in the sheath formation, i.e., relatively close to the Sun. The relatively large distance to the shock also implies that the particle enhancement is not likely a result of a direct magnetic connection to the shock. The expected acceleration mechanism could therefore have been (betatron) acceleration within the contracting mini-FR, as it became compressed in the sheath. This process would increase the perpendicular energy of ion and therefore an anisotropy perpendicular to the magnetic field (i.e., an enhancement of particle fluxes at pitch angle $\sim 90^{\circ}$) would be expected. Unfortunately, due to the limited pitch-angle coverage (only pitch angles close to $90^{\circ}$ are covered), we cannot properly investigate  the possible anisotropy. Correction for the Compton-Getting effect shows that for the captured pitch-angle range ($\sim 50^{\circ}-130^{\circ}$) the increase is consistent with isotropy in the fluid frame. However, the anisotropy created earlier in time could also be lost during the propagation of the CME and its sheath in interplanetary space. We also mention that the fact that the spectrograms did not show  velocity dispersion suggests that particles are not injected from a distance along the magnetic field but rather the S/C is crossing flux tubes filled with accelerated ions.}

{The origin of such mini-FRs, with durations ranging from a few minutes to a few hours in the solar wind, is currently not clear, and it is possible that they have several sources. Previously suggested generation mechanisms include both those of a solar and interplanetary origin. Solar origins includes plasma blobs released from the tips of helmet streamers \citep[e.g.,][]{Sheeley10,Rouillard2011,Sanchez2017,Lavraud2020}, narrow and small CMEs \citep[e.g.,][]{Gilbert2001,Zhao2007}, and `fossil relics' from the Sun \citep[e.g.,][]{Borosvky2008}. Interplanetary origins are generally attributed to reconnection in the HCS \citep[e.g.,][]{Eastwood2002,Moldwin2000, Lavraud2020} or solar wind turbulence \citep[e.g.,][]{Zheng2018}. The close proximity of the mini-FR to HCS crossings in the case studied here makes it likely that the mini-FR was a flux rope formed either due to HCS reconnection or streamer blobs.}

{\cite{Feng2011} showed evidence that mini-FRs are bounded by magnetic reconnection exhausts, and suggested that they gradually diminish with increasing heliospheric distance. Statistical investigation by \cite{Murphy2020} using MESSENGER observations between 0.31 and 0.47 AU found that mini-FRs are more abundant closer to the Sun than reported near the Earth's orbit. Mini-FRs have also been identified in significant numbers in PSP observations during the spacecraft's first heliospheric passes \citep{Zhao2020}. Closer to the Sun, mini-FRs may thus play an increasingly important role in the acceleration and modulation of energetic particles. Upcoming observations by Solar Orbiter, BepiColombo and PSP from their near-Sun orbits will lead to a greater understanding of these issues.} 


\section{Conclusion} \label{sec:conclusion}

{We have investigated the structure of a sheath region ahead of a magnetic cloud observed on April, 19-20, 2020 at Solar Orbiter, L1 (Wind/ACE) and BepiColombo, and its connection to energetic ion enhancements. The sheath was preceded by a slowly propagating and weak interplanetary shock. BepiColombo was located $\sim 1$ AU from the Sun, while Solar Orbiter was at 0.8 AU. All spacecraft were almost radially aligned in terms of heliospheric scales (with a separation of $\sim 900 - 1500$ Earth radii in the east-west direction), and observed the same overall features: a shock, a sheath and a well-defined magnetic cloud. However, their separations were large enough to detect differences in smaller-scale sheath structures. Ion enhancements identified at Solar Orbiter and the L1 spacecraft  occurred in different locations within the sheath and were connected to different substructures: at L1, the particle enhancement was associated with a likely folded or wavy HCS recently crossed by the shock, while at Solar Orbiter, the particle enhancement was confined within a mini flux rope embedded deeper in the sheath. In the investigated case we did not observe significant energetic ion increase related to the shock crossing, but only to the substructures mentioned above. These findings, therefore, indicate that CME sheaths and the localised substructures they embed can contribute to the energization of charged particles in interplanetary space. Future observations by Solar Orbiter, Parker Solar Probe and BepiColombo will be highly important for shedding light on these processes at closer distances to the Sun. This work also highlights the importance of multi-spacecraft observations, both at macroscale (fraction of AU) and mesoscale (few hundreds Earth radii) separations.}


\begin{acknowledgements}{The results presented here have been achieved under the framework of the Finnish Centre of Excellence in Research of Sustainable Space (FORESAIL; Academy of Finland grant numbers 312390 and 336809), which we gratefully acknowledge. EK acknowledges the ERC under the European Union's Horizon 2020 Research and Innovation Programme Project 724391 (SolMAG), and EK and SG acknowledge Academy of Finland Project 310445 (SMASH). This study has also been partially funded through the European Union’s Horizon 2020 research and innovation programme under grant agreement No 101004159 (SERPENTINE). We thank the Solar Orbiter EPD team at the University of Kiel, Germany for the development of the EPD data processing and analysis pipeline. RGH and JRP acknowledge financial support from the Spanish MICIU under project
PID2019-104863RB-I00/AEI/10.13039/501100011033. EA would like to acknowledge the financial support by the Academy of Finland (Postdoctoral Grant No 322455) }\end{acknowledgements}

\bibliographystyle{aa} 
\bibliography{SheathSEP_paper_arXiv.bib} 

\begin{thebibliography}{81}
\expandafter\ifx\csname natexlab\endcsname\relax\def\natexlab#1{#1}\fi

\bibitem[{{Afanasiev} {et~al.}(2015){Afanasiev}, {Battarbee}, \&
  {Vainio}}]{Afanasiev2015}
{Afanasiev}, A., {Battarbee}, M., \& {Vainio}, R. 2015, \aap, 584, A81

\bibitem[{{Afanasiev} {et~al.}(2018){Afanasiev}, {Vainio}, {Rouillard},
  {Battarbee}, {Aran}, \& {Zucca}}]{Afanasiev2018}
{Afanasiev}, A., {Vainio}, R., {Rouillard}, A.~P., {et~al.} 2018, \aap, 614, A4

\bibitem[{{Ala-Lahti} {et~al.}(2019){Ala-Lahti}, {Kilpua}, {Sou{\v{c}}ek},
  {Pulkkinen}, \& {Dimmock}}]{Alalahti2019}
{Ala-Lahti}, M., {Kilpua}, E. K.~J., {Sou{\v{c}}ek}, J., {Pulkkinen}, T.~I., \&
  {Dimmock}, A.~P. 2019, J. Geophys. Res.-Space, 124, 3893

\bibitem[{{Ala-Lahti} {et~al.}(2018){Ala-Lahti}, {Kilpua}, {Dimmock}, {Osmane},
  {Pulkkinen}, \& {Sou{\v{c}}ek}}]{Alalahti2018}
{Ala-Lahti}, M.~M., {Kilpua}, E. K.~J., {Dimmock}, A.~P., {et~al.} 2018, Ann.
  Geophys., 36, 793

\bibitem[{{Axford} {et~al.}(1977){Axford}, {Leer}, \& {Skadron}}]{Axford1977}
{Axford}, W.~I., {Leer}, E., \& {Skadron}, G. 1977, in International Cosmic Ray
  Conference, Vol.~11, International Cosmic Ray Conference, 132

\bibitem[{{Bell}(1978)}]{Bell1978}
{Bell}, A.~R. 1978, \mnras, 182, 147

\bibitem[{{Benkhoff} {et~al.}(2010){Benkhoff}, {van Casteren}, {Hayakawa},
  {Fujimoto}, {Laakso}, {Novara}, {Ferri}, {Middleton}, \&
  {Ziethe}}]{Benkhoff2010}
{Benkhoff}, J., {van Casteren}, J., {Hayakawa}, H., {et~al.} 2010, \planss, 58,
  2

\bibitem[{{Borovsky}(2008)}]{Borosvky2008}
{Borovsky}, J.~E. 2008, Journal of Geophysical Research (Space Physics), 113,
  A08110

\bibitem[{{Burlaga} {et~al.}(1981){Burlaga}, {Sittler}, {Mariani}, \&
  {Schwenn}}]{Burlaga1981}
{Burlaga}, L., {Sittler}, E., {Mariani}, F., \& {Schwenn}, R. 1981, \jgr, 86,
  6673

\bibitem[{{Burlaga}(1988)}]{Burlaga1988}
{Burlaga}, L.~F. 1988, \jgr, 93, 7217

\bibitem[{{Crooker} {et~al.}(1993){Crooker}, {Siscoe}, {Shodhan}, {Webb},
  {Gosling}, \& {Smith}}]{Crooker1993}
{Crooker}, N.~U., {Siscoe}, G.~L., {Shodhan}, S., {et~al.} 1993, \jgr, 98, 9371

\bibitem[{{Das} {et~al.}(2011){Das}, {Opher}, {Evans}, {Loesch}, \&
  {Gombosi}}]{Das2011}
{Das}, I., {Opher}, M., {Evans}, R., {Loesch}, C., \& {Gombosi}, T.~I. 2011,
  \apj, 729, 112

\bibitem[{{Desai} \& {Giacalone}(2016)}]{Desai2016}
{Desai}, M. \& {Giacalone}, J. 2016, Living Reviews in Solar Physics, 13, 3

\bibitem[{{Dresing} {et~al.}(2012){Dresing}, {G{\'o}mez-Herrero}, {Klassen},
  {Heber}, {Kartavykh}, \& {Dr{\"o}ge}}]{Dresing2012}
{Dresing}, N., {G{\'o}mez-Herrero}, R., {Klassen}, A., {et~al.} 2012, \solphys,
  281, 281

\bibitem[{{Eastwood} {et~al.}(2002){Eastwood}, {Balogh}, {Dunlop}, \&
  {Smith}}]{Eastwood2002}
{Eastwood}, J.~P., {Balogh}, A., {Dunlop}, M.~W., \& {Smith}, C.~W. 2002,
  Journal of Geophysical Research (Space Physics), 107, 1365

\bibitem[{{Farrugia} {et~al.}(1999){Farrugia}, {Janoo}, {Torbert}, {Quinn},
  {Ogilvie}, {Lepping}, {Fitzenreiter}, {Steinberg}, {Lazarus}, {Lin},
  {Larson}, {Dasso}, {Gratton}, {Lin}, \& {Berdichevsky}}]{Farrugia1999}
{Farrugia}, C.~J., {Janoo}, L.~A., {Torbert}, R.~B., {et~al.} 1999, in American
  Institute of Physics Conference Series, ed. S.~R. {Habbal}, R.~{Esser}, J.~V.
  {Hollweg}, \& P.~A. {Isenberg}, Vol. 471, 745--748

\bibitem[{{Feng} \& {Wang}(2013)}]{Feng2013}
{Feng}, H. \& {Wang}, J. 2013, \aap, 559, A92

\bibitem[{{Feng} {et~al.}(2007){Feng}, {Wu}, \& {Chao}}]{Zhao2007}
{Feng}, H.~Q., {Wu}, D.~J., \& {Chao}, J.~K. 2007, Journal of Geophysical
  Research (Space Physics), 112, A02102

\bibitem[{{Feng} {et~al.}(2011){Feng}, {Wu}, {Wang}, \& {Chao}}]{Feng2011}
{Feng}, H.~Q., {Wu}, D.~J., {Wang}, J.~M., \& {Chao}, J.~W. 2011, \aap, 527,
  A67

\bibitem[{{Forman}(1970)}]{Forman1970}
{Forman}, M.~A. 1970, \planss, 18, 25

\bibitem[{{Giacalone}(2012)}]{Giacalone2012}
{Giacalone}, J. 2012, \apj, 761, 28

\bibitem[{{Giacalone} {et~al.}(2020){Giacalone}, {Mitchell}, {Allen}, {Hill},
  {McNutt}, {Szalay}, {Desai}, {Rouillard}, {Kouloumvakos}, {McComas},
  {Christian}, {Schwadron}, {Wiedenbeck}, {Bale}, {Brown}, {Case}, {Chen},
  {Cohen}, {Joyce}, {Kasper}, {Klein}, {Korreck}, {Larson}, {Livi}, {Leske},
  {MacDowall}, {Matthaeus}, {Mewaldt}, {Nieves-Chinchilla}, {Pulupa}, {Roelof},
  {Stevens}, {Szabo}, \& {Whittlesey}}]{Giacalone2020}
{Giacalone}, J., {Mitchell}, D.~G., {Allen}, R.~C., {et~al.} 2020, \apjs, 246,
  29

\bibitem[{{Gilbert} {et~al.}(2001){Gilbert}, {Serex}, {Holzer}, {MacQueen}, \&
  {McIntosh}}]{Gilbert2001}
{Gilbert}, H.~R., {Serex}, E.~C., {Holzer}, T.~E., {MacQueen}, R.~M., \&
  {McIntosh}, P.~S. 2001, \apj, 550, 1093

\bibitem[{{Glassmeier} {et~al.}(2010){Glassmeier}, {Auster}, {Heyner},
  {Okrafka}, {Carr}, {Berghofer}, {Anderson}, {Balogh}, {Baumjohann},
  {Cargill}, {Christensen}, {Delva}, {Dougherty}, {Forna{\c{c}}on}, {Horbury},
  {Lucek}, {Magnes}, {Mandea}, {Matsuoka}, {Matsushima}, {Motschmann},
  {Nakamura}, {Narita}, {O'Brien}, {Richter}, {Schwingenschuh}, {Shibuya},
  {Slavin}, {Sotin}, {Stoll}, {Tsunakawa}, {Vennerstrom}, {Vogt}, \&
  {Zhang}}]{Glassmeier2010}
{Glassmeier}, K.~H., {Auster}, H.~U., {Heyner}, D., {et~al.} 2010, \planss, 58,
  287

\bibitem[{{Gold} {et~al.}(1998){Gold}, {Krimigis}, {Hawkins}, {Haggerty},
  {Lohr}, {Fiore}, {Armstrong}, {Holland}, \& {Lanzerotti}}]{Gold1998}
{Gold}, R.~E., {Krimigis}, S.~M., {Hawkins}, S.~E., I., {et~al.} 1998, \ssr,
  86, 541

\bibitem[{{Gold} \& {Hoyle}(1960)}]{Gold1960}
{Gold}, T. \& {Hoyle}, F. 1960, \mnras, 120, 89

\bibitem[{{Good} {et~al.}(2020){Good}, {Ala-Lahti}, {Palmerio}, {Kilpua}, \&
  {Osmane}}]{Good2020}
{Good}, S.~W., {Ala-Lahti}, M., {Palmerio}, E., {Kilpua}, E.~K.~J., \&
  {Osmane}, A. 2020, \apj, 893, 110

\bibitem[{{Gosling} {et~al.}(1987){Gosling}, {Baker}, {Bame}, {Feldman},
  {Zwickl}, \& {Smith}}]{Gosling1987a}
{Gosling}, J.~T., {Baker}, D.~N., {Bame}, S.~J., {et~al.} 1987, \jgr, 92, 8519

\bibitem[{{Gosling} \& {McComas}(1987)}]{Gosling1987b}
{Gosling}, J.~T. \& {McComas}, D.~J. 1987, Geophys. Res. Lett., 14, 355

\bibitem[{{Gosling} {et~al.}(2006){Gosling}, {McComas}, {Skoug}, \&
  {Smith}}]{Gosling2006}
{Gosling}, J.~T., {McComas}, D.~J., {Skoug}, R.~M., \& {Smith}, C.~W. 2006,
  \grl, 33, L17102

\bibitem[{{Horbury} {et~al.}(2020){Horbury}, {O'Brien}, {Carrasco Blazquez},
  {Bendyk}, {Brown}, {Hudson}, {Evans}, {Oddy}, {Carr}, {Beek}, {Cupido},
  {Bhattacharya}, {Dominguez}, {Matthews}, {Myklebust}, {Whiteside}, {Bale},
  {Baumjohann}, {Burgess}, {Carbone}, {Cargill}, {Eastwood}, {Erd{\"o}s},
  {Fletcher}, {Forsyth}, {Giacalone}, {Glassmeier}, {Goldstein}, {Hoeksema},
  {Lockwood}, {Magnes}, {Maksimovic}, {Marsch}, {Matthaeus}, {Murphy},
  {Nakariakov}, {Owen}, {Owens}, {Rodriguez-Pacheco}, {Richter}, {Riley},
  {Russell}, {Schwartz}, {Vainio}, {Velli}, {Vennerstrom}, {Walsh},
  {Wimmer-Schweingruber}, {Zank}, {M{\"u}ller}, {Zouganelis}, \&
  {Walsh}}]{Horbury2020}
{Horbury}, T.~S., {O'Brien}, H., {Carrasco Blazquez}, I., {et~al.} 2020, \aap,
  642, A9

\bibitem[{{Ipavich}(1974)}]{Ipavich1974}
{Ipavich}, F.~M. 1974, \grl, 1, 149

\bibitem[{{Khabarova} {et~al.}(2016){Khabarova}, {Zank}, {Li}, {Maland raki},
  {le Roux}, \& {Webb}}]{Khabarova2016}
{Khabarova}, O.~V., {Zank}, G.~P., {Li}, G., {et~al.} 2016, \apj, 827, 122

\bibitem[{{Kilpua} {et~al.}(2017{\natexlab{a}}){Kilpua}, {Koskinen}, \&
  {Pulkkinen}}]{Kilpua2017}
{Kilpua}, E., {Koskinen}, H.~E.~J., \& {Pulkkinen}, T.~I. 2017{\natexlab{a}},
  Living Reviews in Solar Physics, 14, 5

\bibitem[{{Kilpua} {et~al.}(2017{\natexlab{b}}){Kilpua}, {Balogh}, {von
  Steiger}, \& {Liu}}]{Kilpua2017SSR}
{Kilpua}, E.~K.~J., {Balogh}, A., {von Steiger}, R., \& {Liu}, Y.~D.
  2017{\natexlab{b}}, \ssr, 212, 1271

\bibitem[{{Kilpua} {et~al.}(2020){Kilpua}, {Fontaine}, {Good}, {Ala-Lahti},
  {Osmane}, {Palmerio}, {Yordanova}, {Moissard}, {Hadid}, \&
  {Janvier}}]{Kilpua2020}
{Kilpua}, E. K.~J., {Fontaine}, D., {Good}, S.~W., {et~al.} 2020, Annales
  Geophysicae, 38, 999

\bibitem[{{Kilpua} {et~al.}(2021){Kilpua}, {Good}, {Ala-Lahti}, {Osmane},
  {Fontaine}, {Hadid}, {Janvier}, \& {Yordanova}}]{Kilpua2021}
{Kilpua}, E.~K.~J., {Good}, S.~W., {Ala-Lahti}, M., {et~al.} 2021, Frontiers in
  Astronomy and Space Sciences, 7, 109

\bibitem[{{Kilpua} {et~al.}(2013){Kilpua}, {Isavnin}, {Vourlidas}, {Koskinen},
  \& {Rodriguez}}]{Kilpua2013}
{Kilpua}, E.~K.~J., {Isavnin}, A., {Vourlidas}, A., {Koskinen}, H.~E.~J., \&
  {Rodriguez}, L. 2013, Annales Geophysicae, 31, 1251

\bibitem[{{Kilpua} {et~al.}(2015){Kilpua}, {Lumme}, {Andreeova}, {Isavnin}, \&
  {Koskinen}}]{Kilpua2015}
{Kilpua}, E.~K.~J., {Lumme}, E., {Andreeova}, K., {Isavnin}, A., \& {Koskinen},
  H.~E.~J. 2015, Journal of Geophysical Research (Space Physics), 120, 4112

\bibitem[{{Kouloumvakos} {et~al.}(2019){Kouloumvakos}, {Rouillard}, {Wu},
  {Vainio}, {Vourlidas}, {Plotnikov}, {Afanasiev}, \&
  {{\"O}nel}}]{Kouloumvakos2019}
{Kouloumvakos}, A., {Rouillard}, A.~P., {Wu}, Y., {et~al.} 2019, \apj, 876, 80

\bibitem[{{Krivolutsky} \& {Repnev}(2012)}]{Krivolutsky2012}
{Krivolutsky}, A.~A. \& {Repnev}, A.~I. 2012, Geomagnetism and Aeronomy, 52,
  685

\bibitem[{{Lario} {et~al.}(2019){Lario}, {Berger}, {Decker},
  {Wimmer-Schweingruber}, {Wilson}, {Giacalone}, \& {Roelof}}]{Lario2019}
{Lario}, D., {Berger}, L., {Decker}, R.~B., {et~al.} 2019, \aj, 158, 12

\bibitem[{{Lavraud} {et~al.}(2020){Lavraud}, {Fargette}, {R{\'e}ville},
  {Szabo}, {Huang}, {Rouillard}, {Viall}, {Phan}, {Kasper}, {Bale},
  {Berthomier}, {Bonnell}, {Case}, {Dudok de Wit}, {Eastwood}, {G{\'e}not},
  {Goetz}, {Griton}, {Halekas}, {Harvey}, {Kieokaew}, {Klein}, {Korreck},
  {Kouloumvakos}, {Larson}, {Lavarra}, {Livi}, {Louarn}, {MacDowall},
  {Maksimovic}, {Malaspina}, {Nieves-Chinchilla}, {Pinto}, {Poirier}, {Pulupa},
  {Raouafi}, {Stevens}, {Toledo-Redondo}, \& {Whittlesey}}]{Lavraud2020}
{Lavraud}, B., {Fargette}, N., {R{\'e}ville}, V., {et~al.} 2020, \apjl, 894,
  L19

\bibitem[{{Lavraud} {et~al.}(2009){Lavraud}, {Gosling}, {Rouillard}, {Fedorov},
  {Opitz}, {Sauvaud}, {Foullon}, {Dandouras}, {G{\'e}not}, {Jacquey}, {Louarn},
  {Mazelle}, {Penou}, {Phan}, {Larson}, {Luhmann}, {Schroeder}, {Skoug},
  {Steinberg}, \& {Russell}}]{Lavraud2009}
{Lavraud}, B., {Gosling}, J.~T., {Rouillard}, A.~P., {et~al.} 2009, \solphys,
  256, 379

\bibitem[{{Lavraud} {et~al.}(2010){Lavraud}, {Opitz}, {Gosling}, {Rouillard},
  {Meziane}, {Sauvaud}, {Fedorov}, {Dandouras}, {G{\'e}not}, {Jacquey},
  {Louarn}, {Mazelle}, {Penou}, {Larson}, {Luhmann}, {Schroeder}, {Jian},
  {Russell}, {Foullon}, {Skoug}, {Steinberg}, {Simunac}, \&
  {Galvin}}]{Lavraud2010}
{Lavraud}, B., {Opitz}, A., {Gosling}, J.~T., {et~al.} 2010, Annales
  Geophysicae, 28, 233

\bibitem[{{le Roux} {et~al.}(2018){le Roux}, {Zank}, \& {Khabarova}}]{Roux2018}
{le Roux}, J.~A., {Zank}, G.~P., \& {Khabarova}, O.~V. 2018, \apj, 864, 158

\bibitem[{{Lee}(1983)}]{Lee1983}
{Lee}, M.~A. 1983, \jgr, 88, 6109

\bibitem[{{Lepping} {et~al.}(1995){Lepping}, {Ac{\~{u}}na}, {Burlaga},
  {Farrell}, {Slavin}, {Schatten}, {Mariani}, {Ness}, {Neubauer}, {Whang},
  {Byrnes}, {Kennon}, {Panetta}, {Scheifele}, \& {Worley}}]{Lepping1995}
{Lepping}, R.~P., {Ac{\~{u}}na}, M.~H., {Burlaga}, L.~F., {et~al.} 1995, Space
  Sci. Rev., 71, 207

\bibitem[{{Lin} {et~al.}(1995){Lin}, {Anderson}, {Ashford}, {Carlson},
  {Curtis}, {Ergun}, {Larson}, {McFadden}, {McCarthy}, {Parks}, {R{\`e}me},
  {Bosqued}, {Coutelier}, {Cotin}, {D'Uston}, {Wenzel}, {Sanderson}, {Henrion},
  {Ronnet}, \& {Paschmann}}]{Lin1995}
{Lin}, R.~P., {Anderson}, K.~A., {Ashford}, S., {et~al.} 1995, \ssr, 71, 125

\bibitem[{{Manchester} {et~al.}(2005){Manchester}, {Gombosi}, {De Zeeuw},
  {Sokolov}, {Roussev}, {Powell}, {K{\'o}ta}, {T{\'o}th}, \&
  {Zurbuchen}}]{Manchester2005}
{Manchester}, W.~B., I., {Gombosi}, T.~I., {De Zeeuw}, D.~L., {et~al.} 2005,
  \apj, 622, 1225

\bibitem[{{McComas} {et~al.}(1988){McComas}, {Gosling}, {Winterhalter}, \&
  {Smith}}]{Comas1988}
{McComas}, D.~J., {Gosling}, J.~T., {Winterhalter}, D., \& {Smith}, E.~J. 1988,
  J. Geophys. Res.-Space, 93, 2519

\bibitem[{{Mistry} {et~al.}(2015{\natexlab{a}}){Mistry}, {Eastwood}, \&
  {Hietala}}]{Mistry2015}
{Mistry}, R., {Eastwood}, J.~P., \& {Hietala}, H. 2015{\natexlab{a}}, Journal
  of Geophysical Research (Space Physics), 120, 30

\bibitem[{{Mistry} {et~al.}(2015{\natexlab{b}}){Mistry}, {Eastwood}, {Phan}, \&
  {Hietala}}]{Mistry2015b}
{Mistry}, R., {Eastwood}, J.~P., {Phan}, T.~D., \& {Hietala}, H.
  2015{\natexlab{b}}, \grl, 42, 10,513

\bibitem[{{Moissard} {et~al.}(2019){Moissard}, {Fontaine}, \&
  {Savoini}}]{Moissard2019}
{Moissard}, C., {Fontaine}, D., \& {Savoini}, P. 2019, Journal of Geophysical
  Research (Space Physics), 124, 8208

\bibitem[{{Moldwin} {et~al.}(2000){Moldwin}, {Ford}, {Lepping}, {Slavin}, \&
  {Szabo}}]{Moldwin2000}
{Moldwin}, M.~B., {Ford}, S., {Lepping}, R., {Slavin}, J., \& {Szabo}, A. 2000,
  Geophys. Res. Lett., 27, 57

\bibitem[{{M{\"u}ller} {et~al.}(2013){M{\"u}ller}, {Marsden}, {St. Cyr}, \&
  {Gilbert}}]{Muller2013}
{M{\"u}ller}, D., {Marsden}, R.~G., {St. Cyr}, O.~C., \& {Gilbert}, H.~R. 2013,
  Sol. Phys., 285, 25

\bibitem[{{Murphy} {et~al.}(2020){Murphy}, {Winslow}, {Schwadron}, {Lugaz},
  {Yu}, {Farrugia}, \& {Niehof}}]{Murphy2020}
{Murphy}, A.~K., {Winslow}, R.~M., {Schwadron}, N.~A., {et~al.} 2020, \apj,
  894, 120

\bibitem[{{Nakanotani} {et~al.}(2020){Nakanotani}, {Zank}, \&
  {Zhao}}]{Nakanotani2020J}
{Nakanotani}, M., {Zank}, G.~P., \& {Zhao}, L. 2020, in Journal of Physics
  Conference Series, Vol. 1620, Journal of Physics Conference Series, 012014

\bibitem[{{Neugebauer} {et~al.}(1993){Neugebauer}, {Clay}, \&
  {Gosling}}]{Neugebauer1993}
{Neugebauer}, M., {Clay}, D.~R., \& {Gosling}, J.~T. 1993, \jgr, 98, 9383

\bibitem[{{Neugebauer} \& {Giacalone}(2015)}]{Neugebauer2015}
{Neugebauer}, M. \& {Giacalone}, J. 2015, Journal of Geophysical Research
  (Space Physics), 120, 8281

\bibitem[{{Ogilvie} {et~al.}(1995){Ogilvie}, {Chornay}, {Fritzenreiter},
  {Hunsaker}, {Keller}, {Lobell}, {Miller}, {Scudder}, {Sittler}, {Torbert},
  {Bodet}, {Needell}, {Lazarus}, {Steinberg}, {Tappan}, {Mavretic}, \&
  {Gergin}}]{Ogilvie1995}
{Ogilvie}, K.~W., {Chornay}, D.~J., {Fritzenreiter}, R.~J., {et~al.} 1995,
  Space Sci. Rev., 71, 55

\bibitem[{{Ogilvie} \& {Desch}(1997)}]{Ogilvie1997}
{Ogilvie}, K.~W. \& {Desch}, M.~D. 1997, Adv. Space Res., 20, 559

\bibitem[{{Reames}(2013)}]{Reames2013}
{Reames}, D.~V. 2013, \ssr, 175, 53

\bibitem[{{Rodr{\'\i}guez-Pacheco} {et~al.}(2020){Rodr{\'\i}guez-Pacheco},
  {Wimmer-Schweingruber}, {Mason}, {Ho}, {S{\'a}nchez-Prieto}, {Prieto},
  {Mart{\'\i}n}, {Seifert}, {Andrews}, {Kulkarni}, {Panitzsch}, {Boden},
  {B{\"o}ttcher}, {Cernuda}, {Elftmann}, {Espinosa Lara}, {G{\'o}mez-Herrero},
  {Terasa}, {Almena}, {Begley}, {B{\"o}hm}, {Blanco}, {Boogaerts}, {Carrasco},
  {Castillo}, {da Silva Fari{\~n}a}, {de Manuel Gonz{\'a}lez}, {Drews},
  {Dupont}, {Eldrum}, {Gordillo}, {Guti{\'e}rrez}, {Haggerty}, {Hayes},
  {Heber}, {Hill}, {J{\"u}ngling}, {Kerem}, {Knierim}, {K{\"o}hler}, {Kolbe},
  {Kulemzin}, {Lario}, {Lees}, {Liang}, {Mart{\'\i}nez Hell{\'\i}n}, {Meziat},
  {Montalvo}, {Nelson}, {Parra}, {Paspirgilis}, {Ravanbakhsh}, {Richards},
  {Rodr{\'\i}guez-Polo}, {Russu}, {S{\'a}nchez}, {Schlemm}, {Schuster},
  {Seimetz}, {Steinhagen}, {Tammen}, {Tyagi}, {Varela}, {Yedla}, {Yu},
  {Agueda}, {Aran}, {Horbury}, {Klecker}, {Klein}, {Kontar}, {Krucker},
  {Maksimovic}, {Malandraki}, {Owen}, {Pacheco}, {Sanahuja}, {Vainio},
  {Connell}, {Dalla}, {Dr{\"o}ge}, {Gevin}, {Gopalswamy}, {Kartavykh},
  {Kudela}, {Limousin}, {Makela}, {Mann}, {{\"O}nel}, {Posner}, {Ryan},
  {Soucek}, {Hofmeister}, {Vilmer}, {Walsh}, {Wang}, {Wiedenbeck}, {Wirth}, \&
  {Zong}}]{Rodriguez2020}
{Rodr{\'\i}guez-Pacheco}, J., {Wimmer-Schweingruber}, R.~F., {Mason}, G.~M.,
  {et~al.} 2020, \aap, 642, A7

\bibitem[{{Rouillard} {et~al.}(2011){Rouillard}, {Sheeley}, {Cooper}, {Davies},
  {Lavraud}, {Kilpua}, {Skoug}, {Steinberg}, {Szabo}, {Opitz}, \&
  {Sauvaud}}]{Rouillard2011}
{Rouillard}, A.~P., {Sheeley}, Jr., N.~R., {Cooper}, T.~J., {et~al.} 2011,
  \apj, 734, 7

\bibitem[{{Sanchez-Diaz} {et~al.}(2017){Sanchez-Diaz}, {Rouillard}, {Davies},
  {Lavraud}, {Pinto}, \& {Kilpua}}]{Sanchez2017}
{Sanchez-Diaz}, E., {Rouillard}, A.~P., {Davies}, J.~A., {et~al.} 2017, \apj,
  851, 32

\bibitem[{{Sandroos} \& {Vainio}(2006)}]{Sandroos2006}
{Sandroos}, A. \& {Vainio}, R. 2006, \aap, 455, 685

\bibitem[{{Sheeley} \& {Rouillard}(2010)}]{Sheeley10}
{Sheeley}, Jr., N.~R. \& {Rouillard}, A.~P. 2010, \apj, 715, 300

\bibitem[{{Shodhan} {et~al.}(2000){Shodhan}, {Crooker}, {Kahler},
  {Fitzenreiter}, {Larson}, {Lepping}, {Siscoe}, \& {Gosling}}]{Shodhan2000}
{Shodhan}, S., {Crooker}, N.~U., {Kahler}, S.~W., {et~al.} 2000, \jgr, 105,
  27261

\bibitem[{{Smith} {et~al.}(1998){Smith}, {L'Heureux}, {Ness}, {Acu{\~n}a},
  {Burlaga}, \& {Scheifele}}]{Smith1998}
{Smith}, C.~W., {L'Heureux}, J., {Ness}, N.~F., {et~al.} 1998, \ssr, 86, 613

\bibitem[{{Smith}(2001)}]{Smith2001}
{Smith}, E.~J. 2001, \jgr, 106, 15819

\bibitem[{{Stone} {et~al.}(1998){Stone}, {Frandsen}, {Mewaldt}, {Christian},
  {Margolies}, {Ormes}, \& {Snow}}]{Stone1998}
{Stone}, E.~C., {Frandsen}, A.~M., {Mewaldt}, R.~A., {et~al.} 1998, \ssr, 86, 1

\bibitem[{{Szabo} {et~al.}(2020){Szabo}, {Larson}, {Whittlesey}, {Stevens},
  {Lavraud}, {Phan}, {Wallace}, {Jones-Mecholsky}, {Arge}, {Badman},
  {Odstrcil}, {Pogorelov}, {Kim}, {Riley}, {Henney}, {Bale}, {Bonnell}, {Case},
  {Dudok de Wit}, {Goetz}, {Harvey}, {Kasper}, {Korreck}, {Koval}, {Livi},
  {MacDowall}, {Malaspina}, \& {Pulupa}}]{Szabo2020}
{Szabo}, A., {Larson}, D., {Whittlesey}, P., {et~al.} 2020, \apjs, 246, 47

\bibitem[{{Tessein} {et~al.}(2016){Tessein}, {Ruffolo}, {Matthaeus}, \&
  {Wan}}]{Tessein2016}
{Tessein}, J.~A., {Ruffolo}, D., {Matthaeus}, W.~H., \& {Wan}, M. 2016, \grl,
  43, 3620

\bibitem[{{Vainio} {et~al.}(2009){Vainio}, {Desorgher}, {Heynderickx},
  {Storini}, {Fl{\"u}ckiger}, {Horne}, {Kovaltsov}, {Kudela}, {Laurenza},
  {McKenna-Lawlor}, {Rothkaehl}, \& {Usoskin}}]{Vainio2009}
{Vainio}, R., {Desorgher}, L., {Heynderickx}, D., {et~al.} 2009, \ssr, 147, 187

\bibitem[{{Vainio} \& {Laitinen}(2007)}]{Vainio2007}
{Vainio}, R. \& {Laitinen}, T. 2007, \apj, 658, 622

\bibitem[{{Vainio} \& {Laitinen}(2008)}]{Vainio2008}
{Vainio}, R. \& {Laitinen}, T. 2008, Journal of Atmospheric and
  Solar-Terrestrial Physics, 70, 467

\bibitem[{{Webb} \& {Howard}(2012)}]{Webb2012}
{Webb}, D.~F. \& {Howard}, T.~A. 2012, Living Reviews in Solar Physics, 9, 3

\bibitem[{{Yu} {et~al.}(2014){Yu}, {Farrugia}, {Lugaz}, {Galvin}, {Kilpua},
  {Kucharek}, {M{\"o}stl}, {Leitner}, {Torbert}, {Simunac}, {Luhmann}, {Szabo},
  {Wilson}, {Ogilvie}, \& {Sauvaud}}]{Yu2014}
{Yu}, W., {Farrugia}, C.~J., {Lugaz}, N., {et~al.} 2014, J. Geophys. Res., 119,
  689

\bibitem[{{Zhao} {et~al.}(2020){Zhao}, {Zank}, {Adhikari}, {Hu}, {Kasper},
  {Bale}, {Korreck}, {Case}, {Stevens}, {Bonnell}, {Dudok de Wit}, {Goetz},
  {Harvey}, {MacDowall}, {Malaspina}, {Pulupa}, {Larson}, {Livi}, {Whittlesey},
  \& {Klein}}]{Zhao2020}
{Zhao}, L.~L., {Zank}, G.~P., {Adhikari}, L., {et~al.} 2020, \apjs, 246, 26

\bibitem[{{Zheng} \& {Hu}(2018)}]{Zheng2018}
{Zheng}, J. \& {Hu}, Q. 2018, \apjl, 852, L23

\end{thebibliography}

\begin{appendix} 

\section{Zoom-in to heliospheric current sheet crossings at L1} \label{app:a}

\begin{figure*}
\begin{multicols}{2}
    \includegraphics[width=0.99\linewidth]{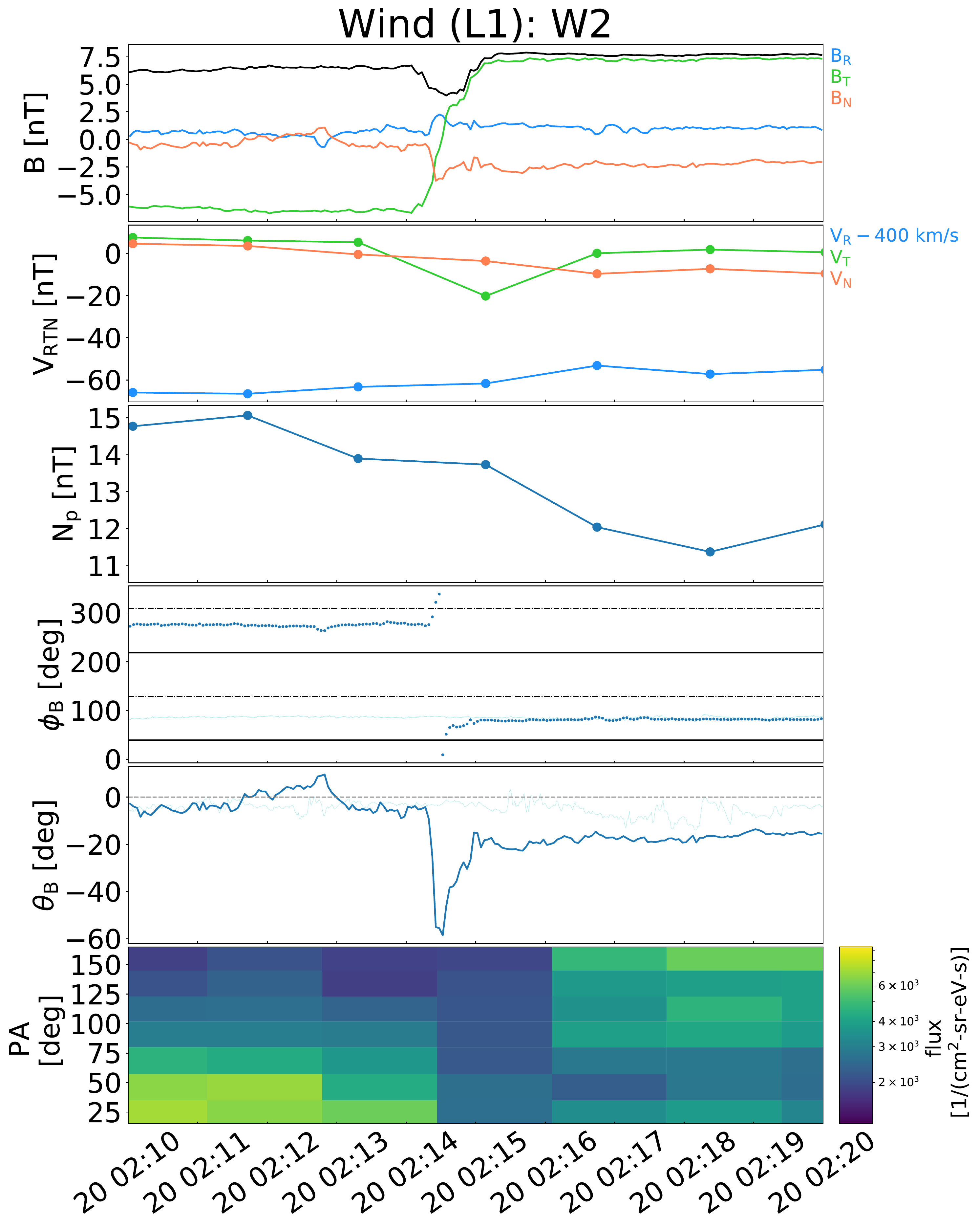}\par 
    \includegraphics[width=0.99\linewidth]{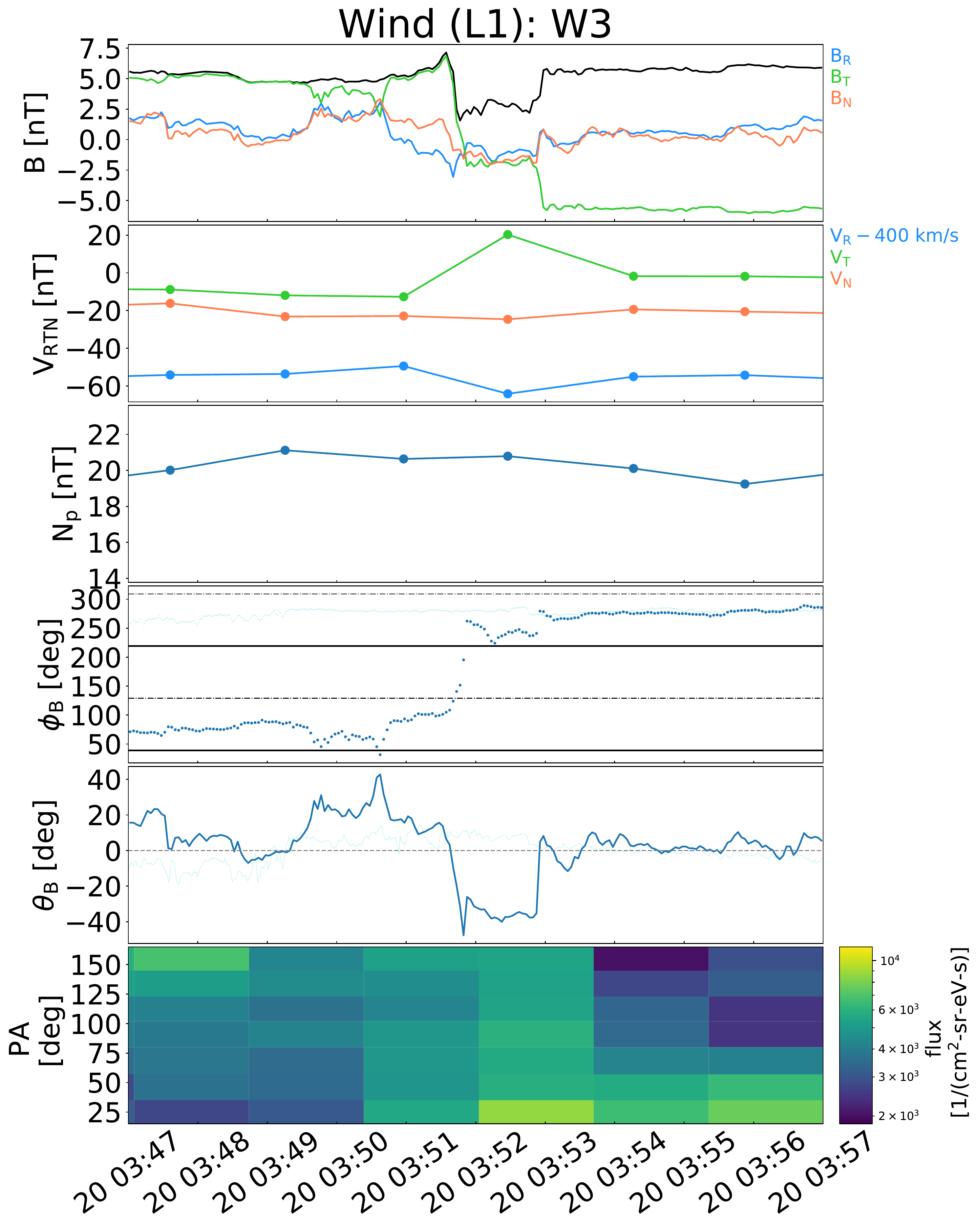}\par 
    \end{multicols}
\begin{multicols}{2}
    \includegraphics[width=0.99\linewidth]{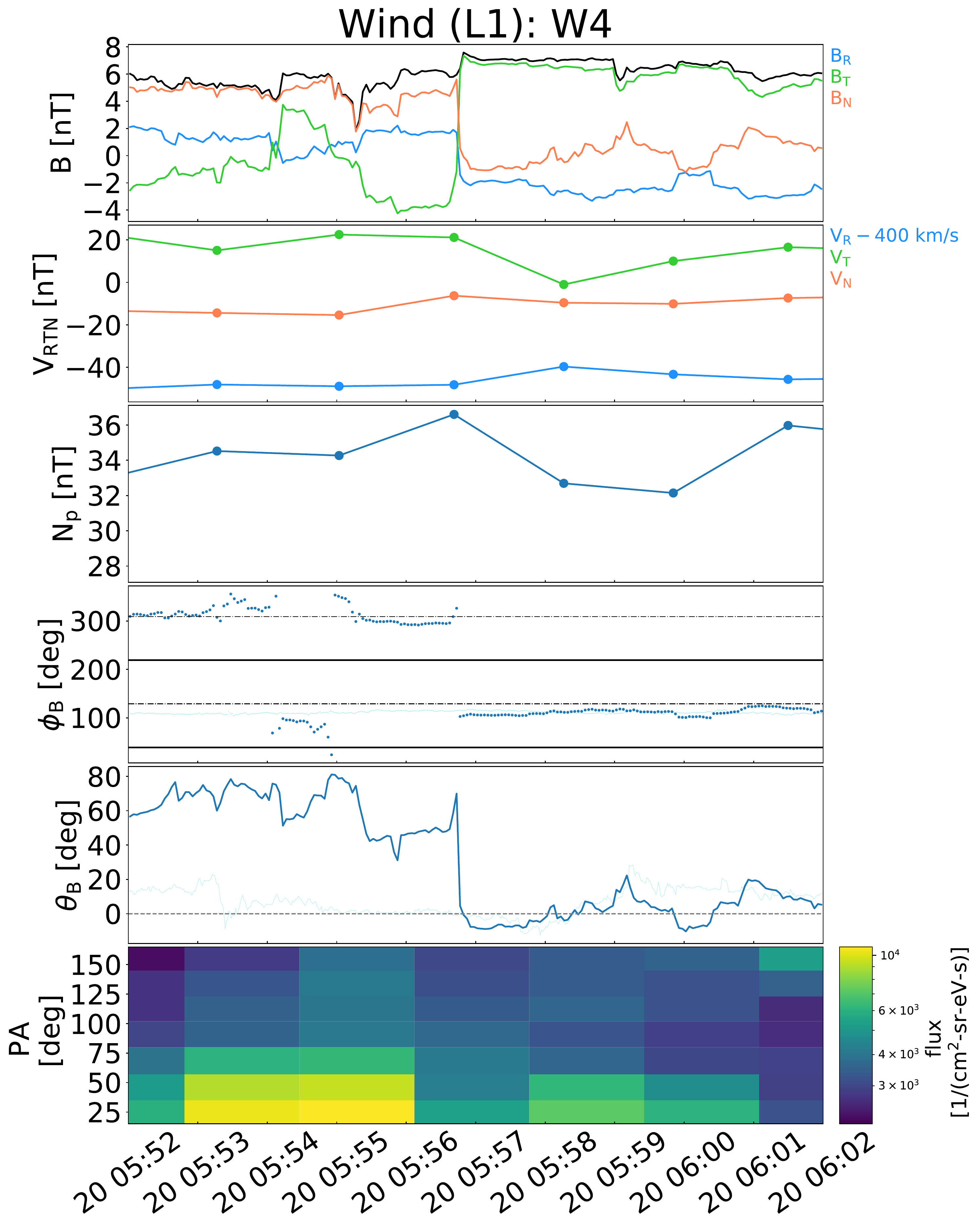}\par
    \includegraphics[width=0.99\linewidth]{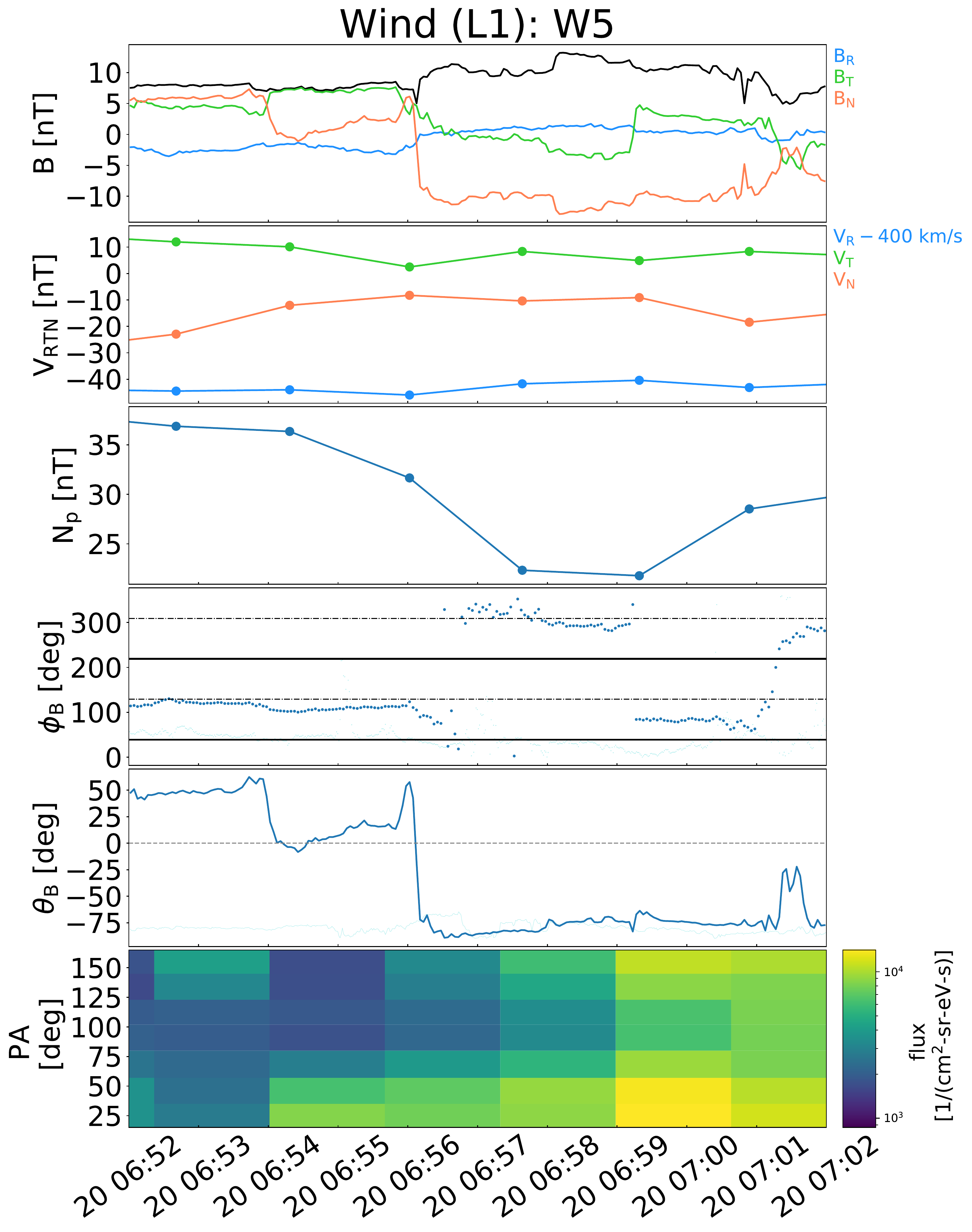}\par
    \end{multicols}
\caption{Zoom-in plots of HCS crossings W2-W5 at Wind, observed within the sheath. The panels show, from top to bottom: The magnetic field magnitude (black) and components in RTN coordinates (blue: $B_R$; green:$B_T$; red: $B_N$), solar wind velocity components in RTN coordinates (blue: $V_R$-400 km/s; green:$V_T$; red: $V_N$), density, magnetic field azimuth and latitude angles (dark blue: Wind, light blue: ACE; the ACE data is not time-shifted), and pitch angle distribution of 265 eV electrons. In the azimuth angle plot, the two dashed-dotted horizontal lines indicate the Parker spiral direction and  solid horizontal lines indicate the division between towards and away sectors.}
\end{figure*}

\section{Compton-Getting transformation} \label{app:b}

At non-relativistic speeds, the particle velocity $\vec{v}'$ in a moving frame (with a relative velocity $\vec{w}$) is connected to the observed velocity in the spacecraft frame $\vec{v}$ by $\vec{v}' = \vec{v} - \vec{w}$ (in the following primed variables denote moving frame quantities whereas unprimed variables refer to the spacecraft frame). Introducing $\theta$, the angle between $\vec{v}$ and $\vec{w}$, one can derive the transformations for the momentum and kinetic energy of the particle, respectively, which are given by 
\begin{eqnarray}
    p' & = & p \sqrt{1-2\frac{w}{v}\cos{\theta}+\frac{w^2}{v^2}} \label{eq:imomentum_transformation}\\
    E' & = & E \cdot \left(1-2\frac{w}{v}\cos{\theta}+\frac{w^2}{v^2} \right) \label{eq:energy_transformation}
\end{eqnarray}

Using the connection of differential intensity, momentum and phase space density, $\frac{dI}{dE} = p^2 \cdot f(\vec{p})$, and the fact that the phase space density is Lorentz-invariant \citep{Forman1970}, we derive the transformation for the differential intensity from the spacecraft frame to a frame moving with a velocity $\vec{w}$:
\begin{eqnarray}
    \left[\frac{dI}{dE} \right]' = \frac{dI}{dE} \cdot \left(1-2\frac{w}{v}\cos{\theta}+\frac{w^2}{v^2} \right) \label{eq:intensity_transformation}
\end{eqnarray}
Also the pitch-angle cosine $\mu$ can be transferred from the spacecraft frame to a moving frame of reference:
\begin{eqnarray}
    \mu'= \frac{\vec{v}'\cdot\vec{B}}{v' B} = \frac{(\vec{v}-\vec{w})\cdot\vec{B}}{v' B}.
\end{eqnarray}

Note that the discussion above assumes that ions are all of the same species (here, protons). In reality, the EPT instrument does not separate the ion species at these energies, so the alpha-particle contribution, if known, should be subtracted from the observed intensities. The alpha to proton ratio can be measured at somewhat higher energies using the SIS instrument on SolO, but at the time of this event, SIS was not yet observing.

\begin{figure*}[ht]
\centering
\includegraphics[width=0.99\linewidth]{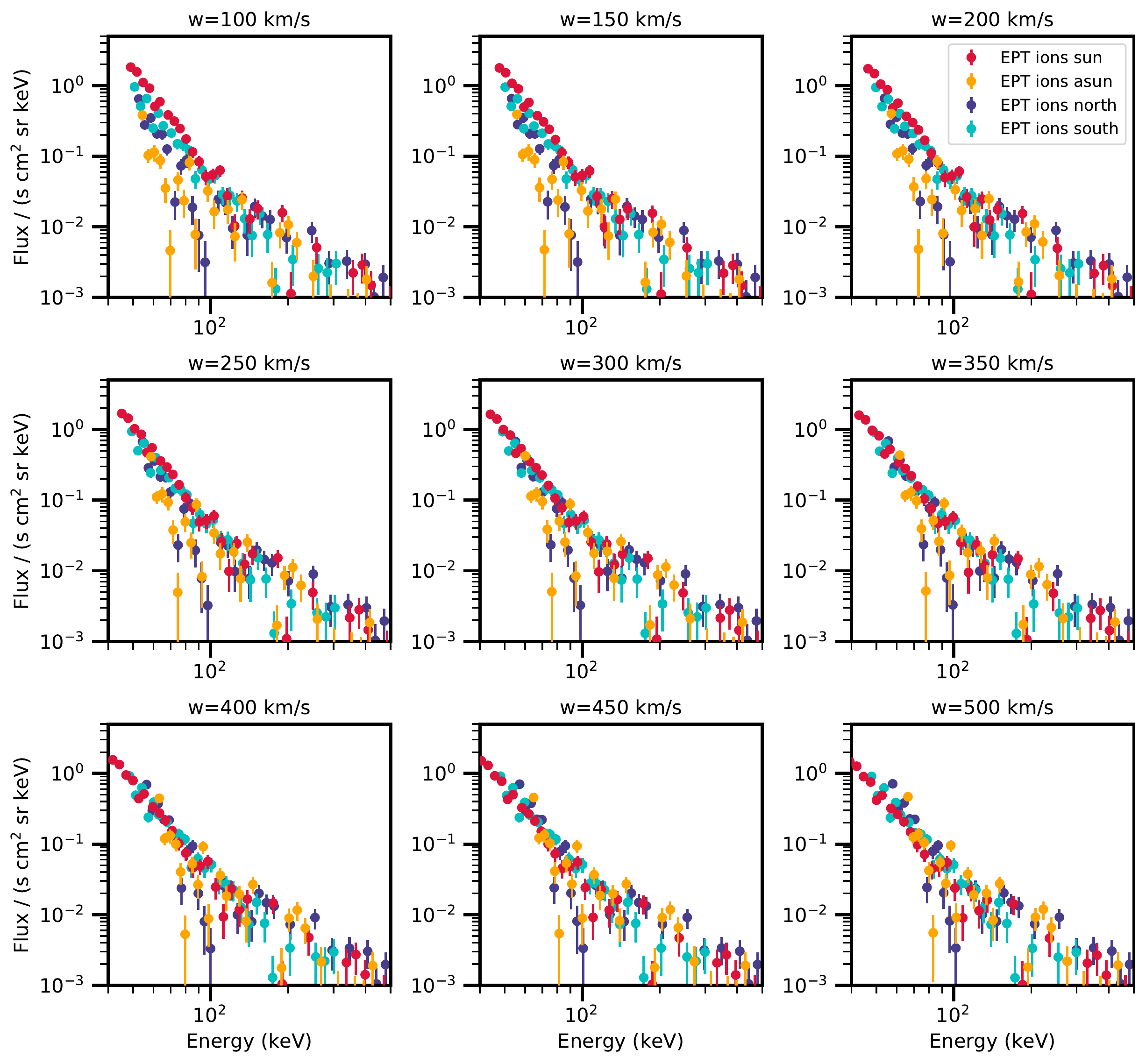}
\caption{Differential energy spectra of ions observed by SolO/EPT for the four viewing directions of EPT, transformed to a moving frame for varying relative velocities. The time interval is 7:00 to 8:00~UT on April 19, 2020, covering the increase of energetic ion fluxes within the mini flux rope.
}
\label{fig:FIG_CG_appendix}
\end{figure*}

Figure~\ref{fig:FIG_CG_appendix} depicts the SolO/EPT differential energy spectra measured in four viewing directions in moving (relative to the S/C) frames of reference for nine values of $w$ in $\vec{w}=w\,\vec{e}_R$. We can see that the value of $w=350$~km/s collapses all the measured spectra, in particular below $\sim$80~keV, quite well into a single trace, indicating that the distribution function is close to isotropic in the solar wind frame. Conversely, assuming that the distribution is isotropic in the wind frame, we can confirm that the wind speed at SolO is close to the propagated value of 350 km/s deduced from Wind observations.

\end{appendix} 

\end{document}